\documentclass[a4paper,11pt]{article}
\usepackage{jheppub}
\usepackage{lineno}
\usepackage{comment}
\usepackage{mathtools}
\usepackage{subcaption}
\usepackage{graphicx}
\usepackage{float}
\usepackage{xcolor}
\usepackage{enumitem}
\usepackage{multirow}

\newcommand*\zb{\bar{z}}

\makeatletter
\def\@fpheader{\relax}
\makeatother

\title{Moments in the CFT Landscape}

\author{Li-Yuan Chiang, David Poland, Gordon Rogelberg}
\affiliation{Department of Physics, Yale University, 217 Prospect St, New Haven, CT 06520, USA}

\emailAdd{li-yuan.chiang@yale.edu}
\emailAdd{david.poland@yale.edu}
\emailAdd{gordon.rogelberg@yale.edu}

\abstract{
	We develop a novel numerical bootstrap for unitary, crossing-symmetric conformal field theories, focusing on moment observables defined as weighted averages over conformal data. Providing a global and coarse-grained probe of the operator spectrum, this framework yields numerically rigorous bounds on the operator distribution using standard semidefinite programming techniques. In the heavy correlator regime, these bounds remain robust and converge rapidly towards analytically-derived power laws. At finite external dimensions, low-lying moments capture corrections to analytic heavy limit results, while reproducing familiar bootstrap solutions such as Ising-model kinks on the boundary of moment space. Most importantly, the moment bootstrap reveals new features in previously unexplored regions of the bootstrap landscape. The lower bounds on moment variables exhibit two continuous families of kinks persisting across $2 < d < 6$, reflecting nontrivial spectral reorganizations connected to underlying operator decoupling phenomena. These results demonstrate that moment variables uncover bootstrap solutions and collective structures that are difficult to access within traditional numerical approaches.
}

\begin{document}

\maketitle

\pagenumbering{roman}
\setcounter{page}{2}
\newpage
\pagenumbering{arabic}
\setcounter{page}{1}

\section{Introduction}

The conformal bootstrap has emerged as one of the most powerful nonperturbative methods for studying conformal field theories (CFT) in general spacetime dimensions. Lying at the core of these techniques are the crossing equations: an infinite family of sum rules that encode the consistency of the operator product expansion (OPE) with conformal symmetry, unitarity, and crossing symmetry. Implemented as a semidefinite program (SDP),  numerically optimized functionals acting on these constraints can be used to make remarkably precise predictions about the low-lying spectra of local operators in a wide range of CFTs such as the 3d Ising CFT~\cite{El-Showk:2012cjh,El-Showk2014-ce,Simmons-Duffin:2016wlq, Kos:2016ysd,Chang:2024whx}, O(N) vector models~\cite{Kos:2013tga,Chester:2019ifh, Liu:2020tpf, Chester:2020iyt}, Gross-Neveu-Yukawa CFTs~\cite{Erramilli2022-yi, Mitchell:2024hix} and their supersymmetric extensions~\cite{Bobev:2015jxa, Bobev:2015vsa, Chester:2015lej, Atanasov:2018kqw, Rong:2019qer, Atanasov:2022bpi}, 3d conformal gauge theories such as quantum electrodynamics with fermionic and scalar matter~\cite{Chester2016-tp, Chester:2017vdh, Li:2020bnb, Li:2021emd, He:2021sto, Albayrak2021-td, He:2021xvg, Chester:2023njo, Chester:2025uxb}, and more. The majority of developments in the bootstrap have been aimed towards the goal of isolating known models in theory space. By using knowledge of low-lying spectra, fusion algebras of local operators, and gaps inspired by equations of motion from Lagrangian descriptions as inputs to the bootstrap equations, we can isolate islands in parameter space that reproduce CFT data, which we can then check against known perturbative results, Monte Carlo, and fuzzy sphere computations. These investigations have since become synonymous with the bootstrap and are widely regarded as the benchmark of its success.

In this paper, we introduce the \emph{numerical moment bootstrap}, a framework in which crossing symmetry and unitarity are used to place numerically rigorous bounds on moments of the OPE data, rather than on individual low-lying operator properties. This provides an alternative way to nonperturbatively map out theory space. We identify moments as natural bootstrap observables that capture global properties of the spectral data appearing in conformal correlators and study these observables with semidefinite programming techniques. This allows us to identify new geometrically privileged points, or \emph{kinks}, in the allowed space of scalar four-point correlators with identical external dimensions. We then characterize these kink structures by identifying unique properties and operator decoupling phenomena within their low-lying spectra. Additionally, we use this new space of bootstrap observables to re-characterize known theories, such as the 3d Ising model. We hope that these results will serve as a starting point for precision investigations of these new bootstrap solutions.

Related observations have appeared in a number of contexts. In~\cite{Paulos:2021jxx} the authors studied bounds on the value of the 4-point function in the crossing-symmetric configuration, observing kinks related to the Ising model. In~\cite{Ghosh:2023onl}, the authors used a basis of analytic functionals dual to extremal solutions for the generalized free field to compute gap-maximizing functionals of a correlator of identical scalar operators in a 2d CFT. In addition to yielding the known kinks associated with $\langle \sigma \sigma\sigma\sigma\rangle$ and $\langle \epsilon \epsilon \epsilon\epsilon \rangle$ in the Ising model, the study also revealed two new kinks that have yet to be fully understood. Similarly, in \cite{Bobev:2015jxa}, the authors uncovered three families of kinks in the space of superconformal field theories (SCFT) in dimensions two through four. One of these kinks was associated with a known family of theories interpolating between the $d=2,\; \mathcal{N} = (2,2)$ minimal model and the theory of a free chiral multiplet in $d=4$. The other two kinks are more mysterious, with one coinciding with a kinematic threshold at which certain operators can appear and the other interpolating to a kink in the space of 4d $\mathcal{N}=1$ SCFTs, which had been previously studied in~\cite{Poland:2011ey, Poland:2015mta, Li:2017ddj}. Similar unexplained kink structures have also been observed in the 4-fermion bootstrap~\cite{Iliesiu:2015qra,Iliesiu:2017nrv,Erramilli:2020rlr}. These investigations not only reveal the uncanny ability of the bootstrap program to isolate known solutions, but also pick out new privileged solutions to crossing that have previously unknown descriptions.

The aforementioned studies make use of functionals that extremize a low-energy property of the OPE spectrum, such as the scaling dimension of the lowest lying non-identity operator (gap-maximization) or the OPE coefficient of a single known operator in the theory (OPE-maximization). These low-lying features of the CFT spectrum can be used to accurately approximate correlation functions with similarly small external scaling dimensions. In \cite{Kraus:2018pax}, the authors showed that the ability to do such an approximation comes from the fact that heavy states with $\Delta > 2\Delta_\phi$ are suppressed in the OPE by factors of size $e^{-(\Delta-2\Delta_\phi)}$, where $\Delta_\phi$ is the external scaling dimension. This fact makes gap-maximizing functionals a useful technique for understanding the space of correlators that are dominated by light states close to the gap.

This leads one to ask how one may understand correlators with large external dimension, where the operators dominating the OPE are no longer close to the gap. The functionals that we analyze in this work are aimed at addressing this problem: they capture \textit{global} properties of the OPE rather than characterizing a single low-lying operator. These functionals, which we refer to as OPE moments, are defined as averages of powers of $\Delta$ and $\ell$ over the OPE sum, weighted by the squared OPE coefficient times a conformal block evaluated at the self-dual point $z=\bar{z}=1/2$. In \cite{Poland:2024hvb}, the authors proved that OPE moments in powers of $\Delta$ enjoy asymptotic two-sided bounds in the limit of large external dimension and are saturated by particular limits of correlators of identical scalars in a generalized free theory.

In this work, we develop and implement the numerical moment bootstrap, using semidefinite programming techniques and the solver \texttt{SDPB}~\cite{Simmons-Duffin:2015qma,Landry:2019qug} to rigorously extend these moment bounds away from the asymptotic regime to finite and small values of the external scaling dimension. This approach not only reproduces the expected large-$\Delta_\phi$ behavior but also resolves the geometry of moment space at finite $\Delta_\phi$, uncovering new families of kinks in the space of CFT correlators across spatial dimensions.

Bounding moments using SDP methods is not new, and has a long history in optimization, probability, and finance, see e.g.~\cite{Nesterov2000, doi:10.1137/S1052623400366802, doi:10.1287/opre.50.2.358.424,doi:10.1137/S1052623401399903,
doi:10.1142/p665}. In physics, similar spaces of moment variables have been utilized recently in e.g.~\cite{Bellazzini:2020cot,Arkani-Hamed:2020blm,Chiang:2021ziz,Bellazzini:2021oaj,Chiang:2022jep,Chiang:2022ltp,Chiang:2023quf} to study the properties of effective field theories (EFTs) in the context of the S-matrix bootstrap. These works make use of the low energy $s\sim0$ expansion of amplitudes of the form
\begin{equation}
	F(s) = \int dM^2 \frac{p(M^2)}{M^2 - s} = \sum_n a_n s^n,
\end{equation}
where $p(M^2)$ is the spectral density and
\begin{equation}
	a_n = \int dM^2 \frac{p(M^2)}{M^2 } \left(\frac{1}{M^2}\right)^n
\end{equation}
denote moments of the positive measure defined by $p(M^2)/M^2$. These moments are linearly related to Wilson coefficients in the effective action of the theory, with the UV cutoff typically set to the mass gap. By the positivity of the spectral density, it follows that the (shifted) Hankel matrices of these moments are positive semidefinite and totally nonnegative:\footnote{This means that the determinants of all minors are nonnegative.}
\begin{equation}
	\begin{bmatrix}
		a_0    & a_1    & a_2    & \ldots \\
		a_1    & a_2    & a_3    & \ldots \\
		a_2    & a_3    & a_4    & \ldots \\
		\vdots & \vdots & \vdots & \ddots
	\end{bmatrix} \geq 0 \quad\text{and} \quad
	\begin{bmatrix}
		a_1    & a_2    & a_3    & \ldots \\
		a_2    & a_3    & a_4    & \ldots \\
		a_3    & a_4    & a_5    & \ldots \\
		\vdots & \vdots & \vdots & \ddots
	\end{bmatrix} \geq 0.
\end{equation}
These conditions restrict the moment sequences associated with amplitudes in a unitary theory to live in a convex ``moment cone.'' In addition to being studied geometrically using the well developed theory of convex polytopes, these moments may be extremized numerically using SDP. These techniques allow one to derive rigorous bounds on all Wilson coefficients in theories admitting a UV completion. With additional assumptions on the system of amplitudes and their relations, one can also isolate known theories such as superstring theory in the space of Wilson coefficients~\cite{Chiang:2023quf}. Related constraints on spectra in the EFT bootstrap have also been derived, leading to nontrivial relations among masses and spins in weakly-coupled theories~\cite{Berman:2024kdh,Berman:2024owc}.

A similar study of moment variables was carried out in the context of the 2d modular bootstrap~\cite{Chiang:2023qgo}, which considered bounds arising from the convex cone of moments of the torus partition function:
\begin{equation}\label{eq:torusmoments}
	Z_{p,q} = \sum_i n_i \Delta_i^p J_i^q.
\end{equation}
In this work, numerical bounds on the leading moments were obtained using SDP methods and were seen to have a significant amount of structure at both small and large values of the central charge, particularly after imposing integrality conditions~\cite{Fitzpatrick:2023lvh,Chiang:2023qgo}.

In the setting of the conformal bootstrap, positive geometry techniques have been previously applied to moments defined as the Taylor coefficients of a 4-point correlation function~\cite{Arkani-Hamed:2018ign,Sen:2019lec,Huang:2019xzm} expanded around the self-dual point:
\begin{equation}
	\mathcal{G}(z) = \sum_n g_n (z-1/2)^n,
\end{equation}
with
\begin{equation}
	g_n = \frac{\mathcal{G}^{(n)}(1/2)}{n!},
\label{eq:taylor}\end{equation}
and $\mathcal{G}^{(n)}(1/2)$ can then be further decomposed into contributions from conformal blocks. Notable results of these studies include a geometric proof of the existence of operators in a window determined by the external dimension~\cite{Sen:2019lec}, and a geometric characterization of extremal functionals~\cite{Huang:2019xzm}.

While this choice of moments is natural from the point of view of the Taylor expansion, they are difficult to interpret and are far removed from the properties of the underlying spectrum. In our work we will focus instead on ``classical" moment variables, taking the form of integer powers of quantum numbers integrated against the positive measure defined by the conformal block decomposition (analogous to~\eqref{eq:torusmoments}), which admits an intuitive statistical interpretation and makes the dependence on the underlying operator spectrum more direct. The Taylor coefficients in~\eqref{eq:taylor} can then be expressed as a linear combination of the moment variables considered in our work, provided one extends the set to include inverse moments that capture the pole structure of the conformal blocks. That said, due to the fact that conformal blocks behave as an approximate power law $\sim r^\Delta$ in the limit of large exchanged dimension, the geometric and classical moments coincide up to a simple prefactor at leading order in the ``heavy" limit of large external dimension. This leads to a number of remarkable properties about geometric moments in the heavy limit, such as their simple description as a Minkowski sum over convex polytopes~\cite{Sen:2019lec}.

There are many ways to functionally decompose a CFT correlator into a positive sum, each of which defines a determinate moment problem. Known decompositions that can be encoded as positive measures include expansions into twist monomials~\cite{vanRees:2024xkb}, radial monomials~\cite{Hogervorst:2013sma,Kravchuk:2020scc}, and conformal blocks. In this work, we study moments defined over a positive measure which encodes the conformal block decomposition of the correlator. The purpose of this choice is that this decomposition converges most rapidly, as infinite families of local operators (which include descendants) are repackaged into a conformal block labeled by the lowest weight primary operator. This allows us to more directly compute moments of light correlators by summing over known data of the low-lying spectrum, weighted by appropriate factors of scaling dimension and spin. In turn, moments of the conformal block decomposition are closely related to previously computed data, unlike those of decompositions that do not distinguish between primary and descendant operators.

This work will be organized as follows: we begin with a review of the moment bootstrap and its numerical implementation in section~\ref{sec:the_moment_bootstrap}. In section~\ref{sec:heavy correlator}, we numerically compute moment bounds for heavy correlators and see agreement with the bounds derived analytically in~\cite{Poland:2024hvb}. We also study the convergence of the bounds and show how one can use the maximum entropy reconstruction method to approximate the OPE density. Then, we discuss moments of the $\langle\sigma\sigma\sigma\sigma\rangle$ correlator in the 3d Ising model in section~\ref{sec:ising}, comparing them against moments computed using its known spectral data. In section~\ref{sec:new-kinks}, we present newly discovered families of kinks in moment space across spatial dimensions, discuss their relation to the locations of known solutions to crossing, such as minimal models, and further draw attention to solutions to crossing that simultaneously extremize a large number of moment variables. Lastly, we conclude with a discussion of our results in section~\ref{sec:discussion} and include additional details in the appendices.

\section{The moment bootstrap}
\label{sec:the_moment_bootstrap}

\subsection{OPE moments}
We begin by defining the moment variables for the conformal block expansion. Consider the four-point correlation function of identical scalar primaries
\begin{equation}
	\langle \phi(x_1) \phi(x_2) \phi(x_3) \phi(x_4) \rangle = \frac{1}{x_{12}^{2\Delta_\phi} x_{34}^{2\Delta_\phi}} G(z, \zb),
\end{equation}
where we have the following conformal block expansion
\begin{equation}
	G(z, \zb) = \sum_{\mathcal{O}} \lambda_{\phi\phi\mathcal{O}}^2 \, g_{\Delta,\ell}(z, \bar{z}).
\end{equation}
We define the moment variables as weighted sums over the conformal data, where the weight is determined by the squared operator product expansion (OPE) coefficients and the conformal blocks evaluated at some fixed kinematics. Specifically, the spectral density is defined as
\begin{equation}\label{def_spec_dens}
	p_{\ell}(\Delta) = \sum_{\mathcal{O}} \lambda_{\phi\phi\mathcal{O}}^2 \, g_{\Delta,\ell}(z, \bar{z}) \, \delta(\Delta - \Delta_{\mathcal{O}}) \delta_{\ell,\ell_{\mathcal{O}}}.
\end{equation}
Since unitarity implies that $\lambda_{\phi\phi\mathcal{O}}^2 \geq 0$, and the conformal blocks are positive at the self-dual point, this defines a positive spectral measure over operator dimensions and spins, leading to the following definition for the moments:
\begin{equation}
	\nu_k = \sum_{\ell} \int d\Delta \, p_\ell(\Delta) \, \Delta^k = \sum_{\mathcal{O}} \lambda_{\phi\phi\mathcal{O}}^2 g_{\Delta_{\mathcal{O}},\ell_{\mathcal{O}}}(z, \zb) \Delta_{\mathcal{O}}^k.
\end{equation}
We can further generalize this idea to moments involving both $\Delta$ and $\ell$:
\begin{equation}\label{unnormalized_moments}
	\nu_{m,n} = \sum_{\ell} \int d\Delta \, p_\ell(\Delta) \, \Delta^m \ell^n = \sum_{\mathcal{O}} \lambda_{\phi\phi\mathcal{O}}^2 g_{\Delta_{\mathcal{O}},\ell_{\mathcal{O}}}(z, \zb) \Delta_{\mathcal{O}}^m \ell_{\mathcal{O}}^n.
\end{equation}
It is immediate from the definition that the moment variables depend on the cross ratios $z$ and $\zb$ through their dependence on the conformal block. In this paper, we focus on moments evaluated at the self-dual point, defined by $z = \bar{z} = 1/2$, where the $s$-$t$ crossing equation is trivially satisfied. All numerical bounds in this paper are computed at this point.

The zeroth moment is simply the correlator evaluated at the chosen kinematic configuration and may possibly be unbounded from above \cite{Paulos:2021jxx} at finite external scaling dimensions, and hence so can the higher moments. Crucially, the \emph{ratios} between moments remain bounded for any finite external dimension \cite{Poland:2024hvb}. Here, we define the \emph{normalized} moments using a normalization with respect to the zeroth moment:
\begin{equation}\label{normalized_moments}
	\langle \Delta^k \rangle \equiv \frac{\nu_k}{\nu_0}, \quad
	\langle \Delta^m \ell^n \rangle \equiv \frac{\nu_{m,n}}{\nu_{0,0}}.
\end{equation}
These moment variables admit a clear statistical interpretation, capturing coarse-grained information about the exchanged conformal data including its mean, variance, skew, kurtosis, etc.

Finally, we may also consider moments defined with the identity excluded. This modification only affects the zeroth moment, shifting it as $\nu_0 \rightarrow \nu_0 - 1$, allowing us to focus more directly on the nontrivial operator content at and beyond the gap state. For convenience, we define
\begin{equation}
	\langle \Delta^k \rangle_* \equiv \frac{\nu_k}{\nu_0 - 1}, \quad
	\langle \Delta^m \ell^n \rangle_* \equiv \frac{\nu_{m,n}}{\nu_{0,0} - 1},
\end{equation}
which will denote the moments with the identity operator excluded from now on.

On a related note, in a previous work \cite{Paulos:2021jxx}, the authors showed that the gap-maximizing solution can be recovered via a simpler optimization problem involving the correlator itself. We find that their observation fits naturally into a broader framework based on moment variables. The value of the correlator at the self-dual point corresponds to the zeroth unnormalized moment, and as we will show, the 3d Ising model not only extremizes the correlator but also exhibits extremal behavior in higher moments, suggesting that the moment bootstrap provides a natural and efficient way to explore the structure of theory space.

\subsection{Numerical bootstrap setup}
\label{numerical_bootstrap_setup}

In this section, we show that the problem of bounding the OPE moments can be understood as an infinite-dimensional linear program, solvable with the conventional extremal functional method via semidefinite programming \cite{Simmons-Duffin:2015qma}.

\paragraph{Crossing equation}
A CFT correlator of identical scalar primaries enjoys the following crossing symmetry equation, which identifies the $s$-channel and the $t$-channel conformal block expansions:
\begin{equation}
	v^{-\Delta_\phi} - u^{-\Delta_\phi} = \sum_\mathcal{O} \lambda_{\phi\phi\mathcal{O}}^2 \left[ u^{-\Delta_\phi} g_{\Delta, \ell}(u, v) - v^{-\Delta_\phi} g_{\Delta, \ell}(v, u) \right] \equiv \sum_\mathcal{O} \lambda_{\phi\phi\mathcal{O}}^2 F_{\Delta, \ell}(u, v),
\end{equation}
where the left-hand side of the equation represents the vacuum contribution. Now, let us rewrite the crossing equation by pulling out the piece that contributes to the spectral density \eqref{def_spec_dens}
\begin{equation}
	v^{-\Delta_\phi} - u^{-\Delta_\phi} = \sum_{\mathcal{O}} \lambda_{\phi\phi\mathcal{O}}^2 g_{\Delta,\ell}(z^*, \zb^*) \frac{F_{\Delta, \ell}(u, v)}{g_{\Delta, \ell}(z^*, \zb^*)},
\end{equation}
where $z^*$ and $\zb^*$ define the point of expansion and are set to $1/2$ in this work. To proceed, we make use of the standard rational approximation of conformal blocks, which allows derivatives of blocks evaluated at the crossing-symmetric point to be expressed as rational functions of the scaling dimension $\Delta$ of the form
\begin{equation}\label{cb_rat_approx}
	\partial_z^m \partial_{\bar{z}}^n g_{\Delta, \ell}(1/2,1/2) \approx \frac{(4r_*)^\Delta}{\prod_{A}(\Delta - \Delta_A^*)} P^{m,n}_\ell(\Delta),
\end{equation}
where $P^{m,n}_\ell(\Delta)$ are polynomials in $\Delta$. We summarize the construction and conventions of the rational approximation in appendix~\ref{appendix:rational_approx}. Here we only emphasize that the approximation is controlled by two parameters, $r_{\max}$ and $\kappa$, which respectively determine the truncation order of the radial expansion used to compute the conformal blocks and the subset of poles retained in the rational function.\footnote{Approximations of this form can be further optimized using the interpolation scheme developed in~\cite{Chang:2025mwt}.} We then act with the derivative operator $\partial_z^m \partial_{\zb}^n$ on both sides to get
\begin{equation}\label{rat_approx_crossing}
	b^{m, n} = \sum_{\mathcal{O}} \lambda_{\phi\phi\mathcal{O}}^2 g_{\Delta,\ell}(z^*, \zb^*) \frac{F^{m,n}_{\ell}(\Delta)}{P_{\ell}^{0,0}(\Delta)} = \sum_{\ell} \int d\Delta p_{\ell}(\Delta) \frac{F^{m,n}_{\ell}(\Delta)}{P_{\ell}^{0,0}(\Delta)} ,
\end{equation}
where $b^{m,n}$ comes from the vacuum contribution, and the null constraint polynomial $F_{\ell}^{m,n} (\Delta)$ comes from expanding the crossing vector $F_{\Delta,\ell}(u,v)$, expressible in terms of a linear combination of $P_{\ell}^{a,b} (\Delta)$ with different $a$ and $b$, whose coefficients are themselves polynomials in $\Delta_\phi$.

\paragraph{Constraints from unitarity}
Unitarity implies that all OPE coefficients \(\lambda_{\phi\phi\mathcal{O}}\) are real, and hence
\begin{equation}
	\lambda_{\phi\phi\mathcal{O}}^2 \geq 0.
\end{equation}
In a self-dual kinematic configuration, the conformal blocks are also non-negative, which guarantees a positive measure. Moreover, unitarity imposes the following lower bounds on the scaling dimension \(\Delta\) of a primary operator with spin \(\ell\):
\begin{equation}
	\Delta \;\geq\;
	\begin{cases}
		\displaystyle \frac{d-2}{2}, & \ell = 0, \\[8pt]
		\displaystyle \ell + d - 2,  & \ell > 0.
	\end{cases}
\end{equation}

\paragraph{Bounding unnormalized moments}
We now ask how small and how large the moment variables can be. Mathematically, this amounts to extremizing a moment subject to the constraints of crossing symmetry and unitarity. Let us write the rational functions in \eqref{rat_approx_crossing} as $N_{\ell}^{m,n}(\Delta) = F_{\ell}^{m,n}(\Delta)/P_{\ell}^{0,0}(\Delta)$, then the problem of maximizing an arbitrary linear combination of the unnormalized moments can be written as
\begin{align}
	\text{Maximize} \quad   & \nu = \sum_{\ell} \int_{\Delta_{\text{min}}(\ell)}^{\infty} d\Delta \, p_{\ell}(\Delta)V_{\ell}(\Delta)\nonumber                                           \\
	\text{subject to} \quad & \sum_{\ell} \int_{\Delta_{\text{min}}(\ell)}^{\infty} d\Delta \, p_{\ell}(\Delta) N^{(\alpha)}_{\ell}(\Delta) = b^{(\alpha)}, \quad p_\ell(\Delta) \geq 0,
\end{align}
where the function $V_{\ell}(\Delta)$ defines the objective function, which is for example chosen to be $\Delta^m \ell^n$ in order to bound $\nu_{m,n}$ as defined in \eqref{unnormalized_moments}. We use the label $\alpha$ as a shorthand for those $(m,n)$ pairs that produce independent crossing constraints.

One may regard the problem as an infinite-dimensional linear program, with a variable $p_{\ell}(\Delta)$ for each physical operator subject to the unitarity constraint. To transform the problem into a form solvable by \texttt{SDPB}, we consider its dual program
\begin{align}
	\text{Minimize} \quad   & \vec{b} \cdot \vec{z} \nonumber \\ 
	\text{subject to} \quad & \vec{z}\cdot\vec{N}_{\ell}(\Delta) - V_{\ell}(\Delta)\geq 0\quad \forall \Delta \geq \Delta_{\text{min}}(\ell) \text{ and } \ell \in \mathcal{S},
	\label{dual_ineq_unnorm_moment}
\end{align}
where the vector dot product denotes a contraction of the index $\alpha$, and $\mathcal{S}$ the included set of spins of the exchanged operators. For identical external scalars, it is the collection of non-negative even numbers (prior to any truncation). One can easily check, by integrating the inequality constraint \eqref{dual_ineq_unnorm_moment} against the positive spectral density, that any dual feasible solution $(y,\vec{z})$ provides a rigorous upper bound on the moment $\nu$:
\begin{equation}
	\nu = \sum_{\ell\in \mathcal{S}}\int_{\Delta_{\text{min}}(\ell)}^{\infty} d\Delta \, p_{\ell}(\Delta)V_{\ell}(\Delta)\nonumber \leq \vec{z} \cdot \sum_{\ell \in\mathcal{S}}\int_{\Delta_{\text{min}}(\ell)}^{\infty} d\Delta \, p_{\ell}(\Delta) \vec{N}_{\ell}(\Delta) = \vec{b} \cdot \vec{z}.
\end{equation}
Therefore, by minimizing $\vec{b} \cdot \vec{z}$ one obtains the strongest upper bound on $\nu$. In practice, one needs to truncate the set $\mathcal{S}$ to a finite size by imposing a spin truncation, $\ell_{\text{max}}$, and consider the crossing constraints up to a finite derivative order $m + n \leq \Lambda$. Imposing a gap assumption corresponds to adjusting $\Delta_{\text{min}}$.

Finally, we note that one can also bound moments defined as weighted averages over operators of a fixed spin, $\ell^*$. This amounts simply to modifying the function $V_{\ell}(\Delta)$ to $V_{\ell}(\Delta) \delta_{\ell,\ell^*}$, and the polynomial program bounding the moment follows straightforwardly:
\begin{align}
	\text{Minimize} \quad   & \vec{b} \cdot \vec{z} \nonumber                                                                                                                                          \\
	\text{subject to} \quad & \vec{z}\cdot\vec{N}_{\ell}(\Delta) - V_{\ell}(\Delta) \delta_{\ell, \ell^*} \geq 0\quad \forall \Delta \geq \Delta_{\text{min}}(\ell) \text{ and } \ell \in \mathcal{S}.
	\label{dual_ineq_unnorm_spin_moment}
\end{align}
We will denote these fixed-spin moments by
\begin{equation}
	\nu_{k}^{(\ell^*)} = \int_{\Delta_{\min}(\ell^*)}^{\infty} d \Delta \, p_{\ell^*}(\Delta) V_{\ell^*}(\Delta).
\end{equation}

\paragraph{Bounding normalized moments}
Our goal is to bound the ratios of the moment variables by solving optimization problems of the form
\begin{align}
	\text{Minimize} \quad   & \frac{\sum_{\ell}\int_{\Delta_{\text{min}(\ell)}}^{\infty} d\Delta \, p_{\ell}(\Delta)V^{(1)}_{\ell}(\Delta) + \beta}{\sum_{\ell}\int_{\Delta_{\text{min}}(\ell)}^{\infty} d\Delta \, p_{\ell}(\Delta)V^{(0)}_{\ell}(\Delta) + \alpha} \nonumber \\
	\text{subject to} \quad & \sum_{\ell}\int_{\Delta_{\text{min}}(\ell)}^{\infty} d\Delta \, p_{\ell}(\Delta) {N}^{(\alpha)}_{\ell}(\Delta) = {b}^{(\alpha)}.
\end{align}
Here we require that the denominator be positive and include the constants \(\alpha\) and \(\beta\) to account for any possible vacuum contribution. For example, setting \(\alpha=\beta=0\) and \(V^{(0)}_{\ell}(\Delta)=1\) reproduces the normalized moments of \eqref{normalized_moments} without the identity operator. Now, we introduce the auxiliary variables
\begin{equation}\label{aux_var}
	\tilde{p}_{\ell}(\Delta) \equiv \frac{p_{\ell}(\Delta)}{\sum_{\ell}\int_{\Delta_{\text{min}}(\ell)}^{\infty} d\Delta' \, p_{\ell}(\Delta')V^{(0)}_{\ell}(\Delta') + \alpha}, \quad t \equiv \frac{1}{\sum_{\ell}\int_{\Delta_{\text{min}}(\ell)}^{\infty} d\Delta' \, p_{\ell}(\Delta')V^{(0)}_{\ell}(\Delta') + \alpha},
\end{equation}
so that the problem becomes a standard linear program:
\begin{align}\label{linear_frac}
	\text{Minimize} \quad   & \sum_{\ell\in\mathcal{S}}\int_{\Delta_{\text{min}}(\ell)}^{\infty} d\Delta \, \tilde{p}_{\ell}(\Delta)V^{(1)}_{\ell}(\Delta) + \beta \, t \nonumber                                                    \\
	\text{subject to} \quad & \sum_{\ell\in \mathcal{S}}\int_{\Delta_{\text{min}}(\ell)}^{\infty} d\Delta \, \tilde{p}_{\ell}(\Delta) \vec{N}_{\ell}(\Delta) = t \, \vec{b}, \nonumber                                               \\
	                        & \sum_{\ell\in \mathcal{S}}\int_{\Delta_{\text{min}}(\ell)}^{\infty} d\Delta \, \tilde{p}_{\ell}(\Delta) V_{\ell}^{(0)}(\Delta) + \alpha \, t = 1, \quad \tilde{p}_\ell(\Delta) \geq 0, \quad t \geq 0,
\end{align}
whose dual is
\begin{align}
	\text{Maximize} \quad   & -y \nonumber                                                                                                                                                                                    \\
	\text{subject to} \quad & V^{(1)}_{\ell}(\Delta) + y V^{(0)}_{\ell}(\Delta) + \vec{z}\cdot\vec{N}_{\ell}(\Delta)  \geq 0 \quad \forall \Delta \geq \Delta_{\text{min}}(\ell) \text{ and } \ell \in \mathcal{S}, \nonumber \\
	                        & \beta + \alpha y - \vec{b} \cdot \vec{z} \geq 0,
\end{align}
which is a polynomial program that can be solved using \texttt{SDPB}.

\paragraph{Spectrum extraction}
Having cast the problem of bounding the moments into an \texttt{SDPB}-solvable form, we now turn to spectrum extraction.  At the optimal solution, the extremal spectrum follows from the complementary slackness conditions. This procedure is described in detail in \cite{Simmons-Duffin:2016wlq} and is implemented in the built-in \texttt{spectrum} program of \texttt{SDPB}. For unnormalized moments, \texttt{spectrum} outputs the scaling dimensions, spins, and the squared OPE coefficients (up to the prefactor appearing in the rational approximation \eqref{cb_rat_approx}). From these, one directly reconstructs the spectral density defined in \eqref{def_spec_dens}. For normalized moments, \texttt{spectrum} instead returns the normalized probabilities \(\tilde{p}_\ell(\Delta)\) introduced in \eqref{aux_var}. To recover the actual spectral density, one multiplies \(\tilde{p}_\ell(\Delta)\) by the value of the correlator, \(t^{-1}\), which can be fixed by enforcing any one of the null constraints in \eqref{linear_frac}. The normalized OPE coefficients are defined in a similar way:
\begin{equation}
	\tilde\lambda_{\phi\phi\mathcal{O}} = \sqrt{t}\lambda_{\phi\phi\mathcal{O}}.
\end{equation}

\subsection{Summary of results}

\begin{figure}[tbp]
	\centering
	\begin{subfigure}[t]{0.97\textwidth}
		\centering
		\caption{$d = 2$}
		\includegraphics[width=0.85\textwidth]{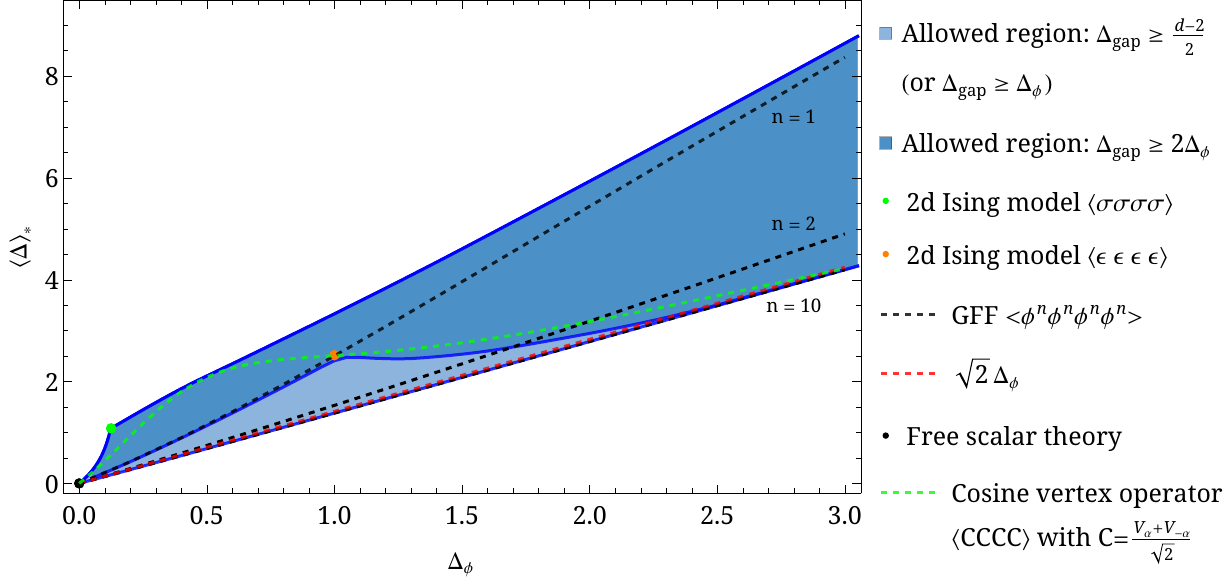}
		\label{fig:summary_2d}
	\end{subfigure}
	\begin{subfigure}[t]{0.97\textwidth}
		\centering
		\caption{$d = 3$}
		\includegraphics[width=0.85\textwidth]{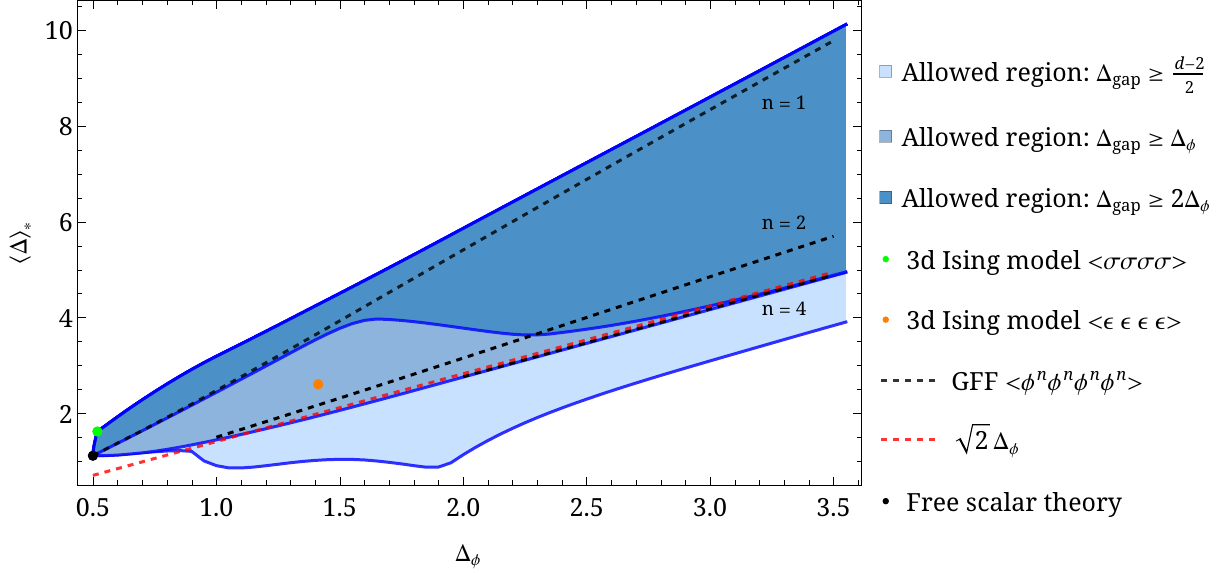}
		\label{fig:summary_3d}
	\end{subfigure}
	\begin{subfigure}[t]{0.97\textwidth}
		\centering
		\caption{$d = 4$}
		\includegraphics[width=0.85\textwidth]{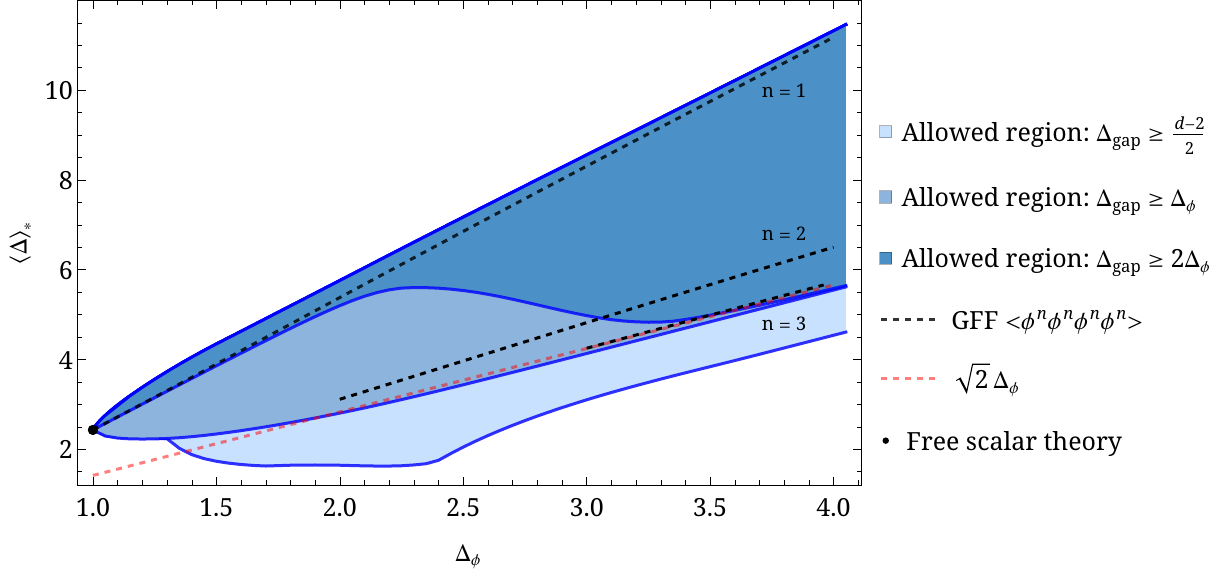}
		\label{fig:summary_4d}
	\end{subfigure}
	\caption{Bounds on the first normalized moment (excluding the identity) as a function of the external scaling dimension, computed at $\Lambda = 23$ (see appendix~\ref{appendix:rational_approx} for other truncation parameters). The shaded regions show the allowed moment space under different gap assumptions, with darker shades indicating stronger gaps. The solid points correspond to moments extracted from $\langle\sigma\sigma\sigma\sigma\rangle$ and $\langle\epsilon\epsilon\epsilon\epsilon\rangle$ in the Ising model, while the black dashed lines indicate GFF solutions. The upper bounds are insensitive to the gap choice. See appendix~\ref{appendix:minimal_model} for details on the minimal model and vertex operator moment computations.}
	\label{fig:summary_all}
\end{figure}

Figures~\ref{fig:summary_2d}, \ref{fig:summary_3d}, and \ref{fig:summary_4d} present our main numerical results, showing the allowed regions for the first normalized moment variable,
\begin{equation}
	\langle\Delta\rangle_* = \frac{\nu_1}{\nu_0 - 1},
\end{equation}
defined excluding the identity operator, across various spatial dimensions. The geometry of the moment variable exhibits a remarkably rich structure, where both the upper and lower bounds display pronounced nonlinear features, showing that the space of consistent moment values is far from trivial.

The dashed curves in the plots indicate the moments computed from generalized free field (GFF) correlators. As the external dimension $\Delta_\phi$ increases, the numerical bounds rapidly converge towards linear trajectories, which coincide with the GFF correlators at leading order. Section \ref{sec:heavy correlator} analyzes this behavior in the large-$\Delta_\phi$ limit and demonstrates agreement between the numerical bounds and analytical results derived in the heavy limit.

In two and three dimensions, sharp kinks are clearly visible in the upper bounds, located precisely at the external dimensions corresponding to the critical Ising model. Section \ref{sec:ising} provides a detailed study of these points, including the leading moments and the spectral interpretation of these kinks that characterize the Ising CFTs.

Finally, in the lower bounds, we identify sharp turning points that persist across dimensions, forming two new continuous families of kinks! Section~\ref{sec:new-kinks} is devoted to exploring the origin of these features in terms of the corresponding extremal spectra, which includes the decoupling of operators along the boundaries and an analysis of possible connections to the fake-primary effect \cite{Erramilli:2020rlr, Karateev:2019pvw}.

\newpage
\section{Moments in the heavy correlator limit}
\label{sec:heavy correlator}

To demonstrate the power of the numerical moment bootstrap, in this section we will study the heavy correlator limit, a regime in which standard gap-maximizing functionals converge extremely slowly in the numerical bootstrap. In contrast, our numerical moment bounds show excellent agreement with those analytically computed at leading order in large $\Delta_\phi$. We then proceed with a spectral reconstruction of the operator distribution obtained from the numerical moment bounds.

\subsection{Moments of heavy correlators}

In the previous work~\cite{Poland:2024hvb}, the authors derived the following two sided bounds on normalized moments in the limit of large external dimension $\Delta_{\phi}$:
\begin{equation}\label{heavy_moments_power_law_bounds}
	2^{\frac{k}{2}}
	\leq
	\lim_{\Delta_\phi \to\infty}\frac{\langle \Delta^{k} \rangle}{\Delta_\phi^k}
	\leq
	2^{\frac{3k}{2} - 1}.
\end{equation}
The correlators whose moments saturate these bounds are given by four-point functions of normal ordered products of elementary fields $\varphi^n$ in a generalized free theory. The lower bound is saturated by the case where $\Delta_\varphi$ is held finite while $n \to \infty$, resulting in the total external dimension $\Delta_\phi = \Delta_{\varphi^n}$ becoming infinite, while the upper bound is saturated by the case of $\Delta_\varphi \to \infty$ with $n=1$.

In the light correlator regime, the standard numerical bootstrap provides rigorous bounds through SDP, which converge to known solutions such as the critical Ising model. A natural question is: \emph{can the numerical moment bootstrap effectively probe the heavy correlator regime?}

\begin{figure}[tpb]
	\centering
	\begin{subfigure}[t]{\textwidth}
		\centering
		\includegraphics[width=\textwidth]{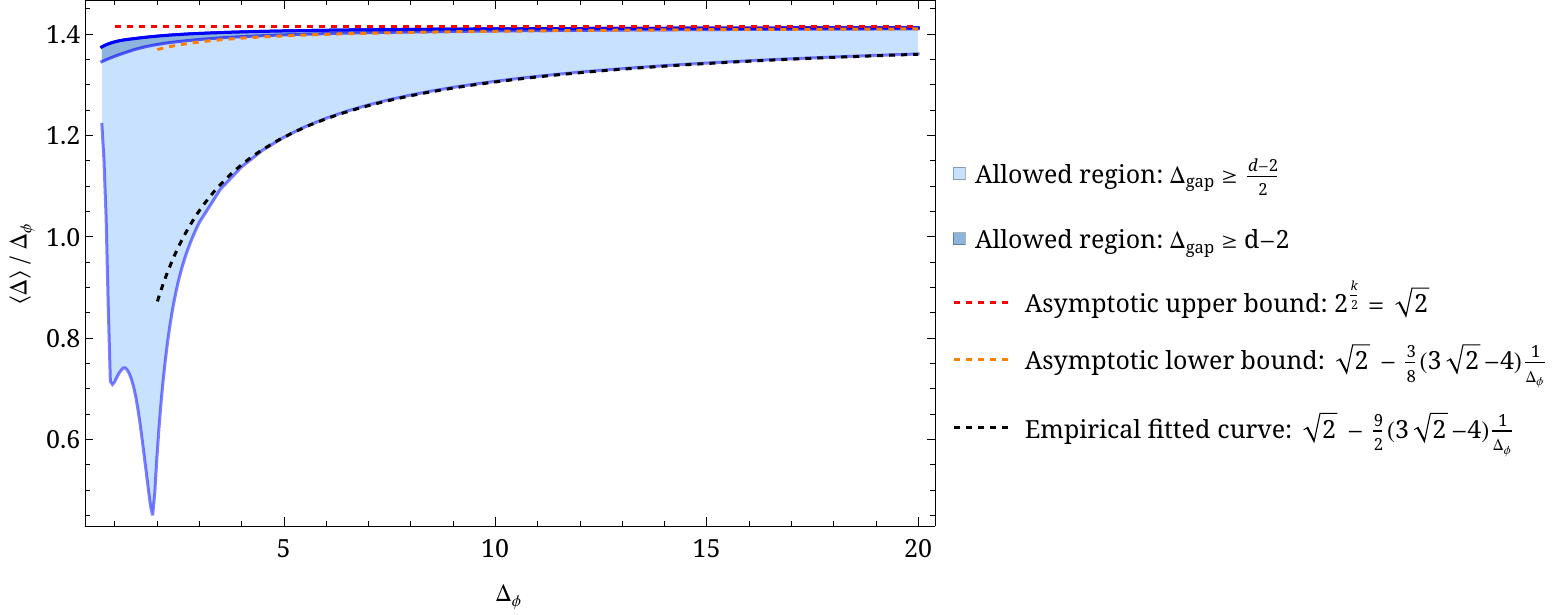}
		\label{fig:heavy1}
	\end{subfigure}
	\begin{subfigure}[t]{\textwidth}
		\centering
		\includegraphics[width=\textwidth]{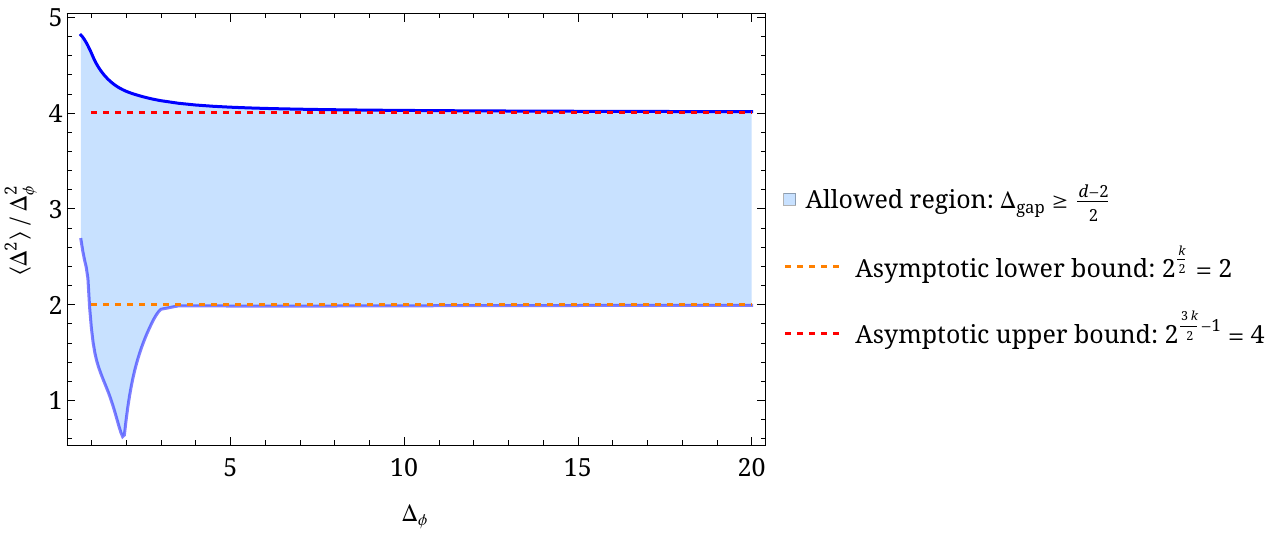}
		\label{fig:heavy2}
	\end{subfigure}
	\begin{subfigure}[t]{\textwidth}
		\centering
		\includegraphics[width=\textwidth]{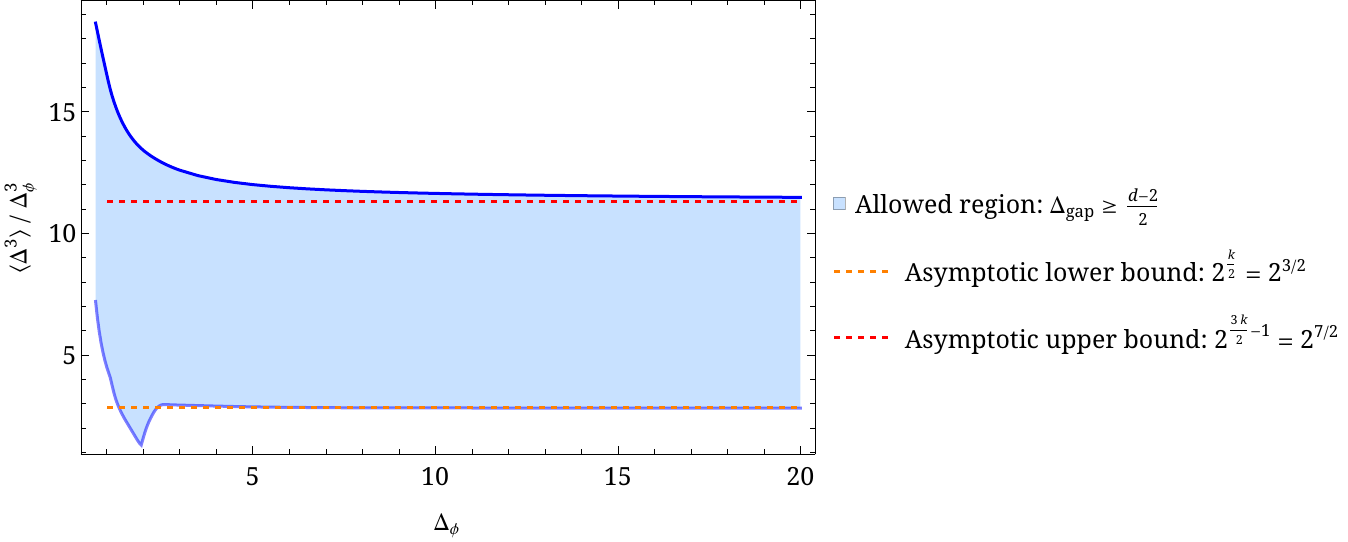}
		\label{fig:heavy3}
	\end{subfigure}
	\caption{Two-sided bounds on the first three normalized moments including the identity operator as a function of the external scaling dimension $\Delta_{\phi}$ in $d = 3$, computed at $\Lambda = 23$. The vertical axes are normalized by $\Delta_{\phi}^k$, so that the asymptotic analytic bounds in~\eqref{heavy_moments_power_law_bounds} appear as straight horizontal lines.}
	\label{fig:heavy moment-bounds}
\end{figure}

It turns out that the numerical moment bootstrap performs strikingly well: the bounds converge rapidly toward the analytic power laws, even deep in the heavy correlator regime. The reason for this is that moment variables are not sensitive to the precise heavy operator spectrum and instead only probe the collective behavior of operator contributions. The fact that $\Lambda$ derivatives acting on the conformal block produce an approximate rational function in $\Delta$, which tends towards a polynomial of degree $\Lambda$ at large values of exchanged dimension, means that to accurately compute the asymptotic behavior of $\langle\Delta^n\rangle$ moments for $n \leq\Lambda$, one needs to only use derivative functionals up to order $\Lambda$, where the precise bounds in the light correlator regime are corrected by including higher derivatives in the functional basis. To quantify this behavior, we examine in figure~\ref{fig:heavy moment-bounds} the leading normalized moments obtained from the numerical bootstrap in spatial dimension $d = 3$. The analytic power-law bounds in~\eqref{heavy_moments_power_law_bounds} are shown as colored dashed lines for comparison. While we numerically bound the moments in $d = 3$, similar behavior can also be observed in other dimensions in the heavy correlator limit.

For the first moment, the asymptotic upper and lower bounds coincide at leading order, both given by $\sqrt{2}\,\Delta_\phi + O(1)$. In this case, one can derive an even stronger lower bound by assuming a scalar gap of $\Delta_{\text{gap}} \geq d - 2$ in $d \leq 4$:
\begin{equation}\label{heavy_lower_subleading}
	\sqrt{2}\,\Delta_{\phi} - \frac{d}{8}(3 \sqrt{2} - 8) \frac{\nu_0 - 1}{\nu_0}
	\leq \langle \Delta \rangle
	\leq \sqrt{2}\,\Delta_{\phi}.
\end{equation}
Under this additional assumption, the numerical lower bounds shrink significantly, leaving a narrow region that approaches the asymptotic curve
\begin{equation}
	\langle \Delta \rangle
	\rightarrow
	\sqrt{2}\,\Delta_{\phi} - \frac{d}{8}(3 \sqrt{2} - 4)
\end{equation}
at large~$\Delta_\phi$, in excellent agreement with the analytic lower bounds.

When instead assuming the most conservative scalar gap from unitarity, i.e., $\Delta_{\text{gap}} \geq (d-2)/2$, the derivation of~\eqref{heavy_lower_subleading} no longer applies. Indeed, the numerical results lie below that curve, reflecting a significant contribution from the lightest scalar operator. Interestingly, the lower bound in $d=3$ can still be captured very well by the empirical fit
\begin{equation}\label{empirical_fit}
	\frac{\langle \Delta \rangle}{\Delta_{\phi}}
	\approx
	\sqrt{2}
	- \frac{9}{2}(3 \sqrt{2} - 4)\frac{1}{\Delta_{\phi}}.
\end{equation}
It would be interesting to derive this analytically. Examining the extremal spectra along the lower boundary reveals an interesting phenomenon: beyond a certain $\Delta_{\phi}$, the lightest scalar operator actually begins to saturate the unitarity bound. We will return to this observation in section~\ref{sec:new-kinks}, where we discuss its connection to the fake-primary effect.

\subsection{Comparison to gap maximization}

\begin{figure}[tbp]
	\centering
	\includegraphics[width=\textwidth]{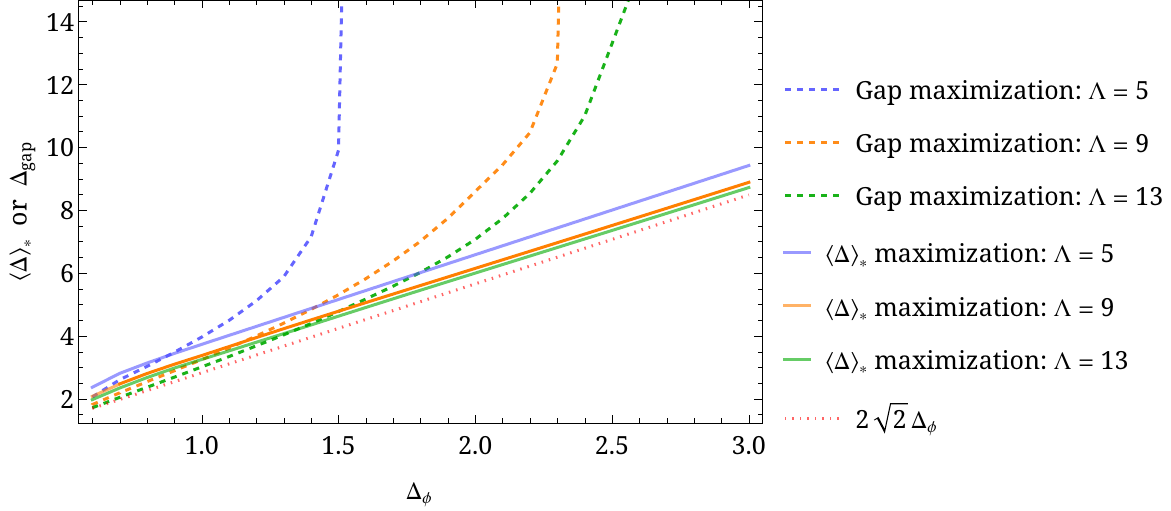}
	\caption{
		Gap maximization compared to moment maximization (without the identity contribution) at the same derivative order. The moment bounds converge rapidly toward the linear trajectory $2 \sqrt{2}\,\Delta_{\phi}$ and reproduce the correct asymptotic scaling at very low derivative order. In contrast, gap maximization converges much more slowly and fails to produce any bound at large external scaling dimension. Although the two procedures optimize different observables, this comparison illustrates that, in the heavy correlator regime, the moments provide more natural observables for the bootstrap. 
	}
	\label{gapMax_vs_moment}
\end{figure}

The numerical moment bootstrap performs remarkably well in the heavy correlator regime. Here we make a brief comparison between maximizing the first normalized moment (with the identity removed), $\langle \Delta \rangle_*$, and the conventional gap-maximization approach. The optimal gaps and moments, obtained at the same derivative orders, are shown in figure~\ref{gapMax_vs_moment}.

As seen in the plot, the optimal gap bounds at different derivative orders converge reasonably well at small external scaling dimension, following an approximately linear trajectory. However, at any fixed derivative order $\Lambda$, the gap-maximization bounds deteriorate as the correlator becomes increasingly heavy: beyond a sufficiently large external weight, the method ceases to produce meaningful bounds.

In contrast, the moment is always bounded even at an extremely low derivative order, such as $\Lambda = 3$,\footnote{While an upper bound on $\langle \Delta \rangle$ can be obtained even at $\Lambda = 1$~\cite{Poland:2024hvb}, it requires a slightly larger $\Lambda$ to obtain an upper bound on $\langle \Delta \rangle_*$.} and converges remarkably rapidly toward the linear trajectory $2 \sqrt{2} \Delta_{\phi}$. This illustrates a key qualitative difference: both conceptually and numerically, moment variables provide a natural and effective language for the conformal bootstrap in the heavy correlator regime. Rather than focusing on individual operators, it is more informative to characterize the average properties of the full CFT spectrum.

\subsection{Max-entropy reconstruction}
Why does the numerical bootstrap perform so well in the heavy correlator regime? A natural first attempt is to examine the extremal spectra directly. However, the extremal spectrum by itself does not provide a meaningful reconstruction of the OPE measure: it only specifies the positions of finitely many operators, whose number can grow at most linearly with $\Lambda$, while correlators at large external weight are governed by the collective contribution of many states. Indeed, when we inspect the extremal spectra extracted from the numerical bounds, the individual operator locations appear scattered, with no obvious Gaussian profile of the kind predicted analytically, which is not surprising. The heavy limit concerns the coarse-grained structure of the OPE measure, and such structure is inherently invisible at the level of individual states.

A key observation, however, comes to the rescue: at each external weight $\Delta_\phi$, the upper bounds of the leading moments are all saturated by the \emph{same} extremal spectrum. This is in sharp contrast to the lower bounds on the leading moments, whose extremal spectra differ almost everywhere. Because the upper bounds at a fixed $\Delta_{\phi}$ share the same spectrum, it becomes meaningful to attempt an inversion: \emph{given the first few optimal moments of a single underlying measure, can we reconstruct the measure itself?}

This leads us to a version of the classical moment problem, which concerns the existence and uniqueness of a positive measure compatible with a given set of moments. In our setting, however, we only have access to a finite number of leading moments. As a result, the reconstruction problem is intrinsically underdetermined: many distinct positive measures are compatible with the same finite set of moments, unless special degeneracies such as a flat extension occur.

This lack of uniqueness, nevertheless, does not imply that a meaningful spectral reconstruction is impossible, but instead reflects the inherently coarse-grained nature of the heavy correlator regime. Motivated by the observation of \cite{Poland:2024hvb} that, in the $\Delta_\phi \to \infty$ limit, double-twist families in GFF correlators approach Gaussian (maximum-entropy) distributions determined by their mean and variance, we adopt the maximum-entropy principle as a physically motivated prescription to select the simplest representative among all measures compatible with the known moments. This approach allows us to systematically study finite-$\Delta_\phi$ corrections to this universal structure. In appendix~\ref{appendix:max_entropy} we describe the details of how this is implemented.

\begin{figure}[tbp]
	\centering
	\includegraphics[width=\textwidth]{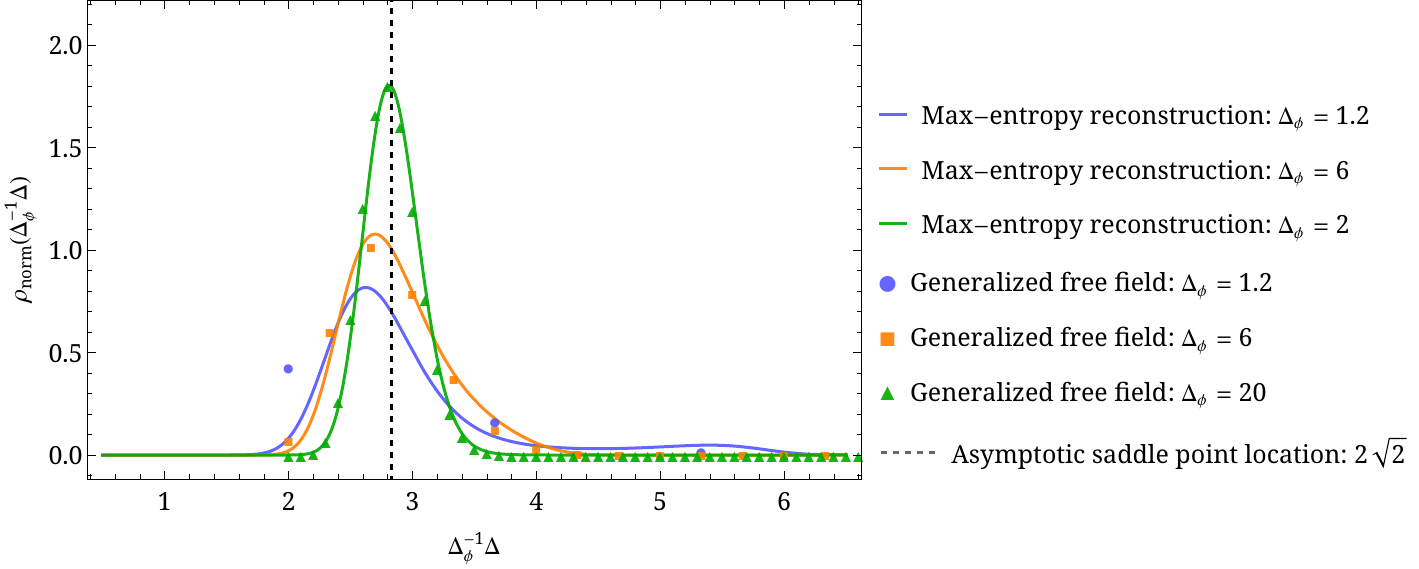}
	\caption{The maximum-entropy reconstruction of the OPE measure (smooth solid curves) compared to the exact discrete measure of the generalized free field correlator $\langle \phi\phi\phi\phi \rangle$. Different colors represent different external scaling dimensions $\Delta_\phi$. The OPE measure is normalized to integrate to unity against $\Delta_{\phi}^{-1} \Delta$.}
	\label{maxEnt}
\end{figure}

Figure~\ref{maxEnt} compares the resulting maximum-entropy reconstruction based on the first five moments with the exact discrete OPE measure of the $\langle \phi\phi\phi\phi\rangle$ generalized free field. The reconstructed density shows clear convergence toward the generalized free field OPE distribution as we approach the heavy limit. At the same time, the numerics reveal systematic deviations from a perfect Gaussian, capturing next-to-leading-order corrections to the upper bounds beyond the leading analytic asymptotic approximation.

From the numerical perspective, as established in section~\ref{sec:the_moment_bootstrap}, the moment variables are specific linear combinations of derivative operators acting on the correlator that yield weighted averages of conformal data. Yet, as demonstrated here, they exhibit behavior fundamentally different from standard functionals that extremize a single operator’s property. The moments effectively ``sense’’ where the dominant collective contribution from a large number of states is concentrated, and reproduce the correct coarse-grained behavior using only a finite number of Dirac delta functions in the extremal spectral density. This remarkable feature makes moment variables not only an efficient tool for discovering light correlator CFTs, but also a powerful probe even in the heavy correlator regime.

Finally, as is visible in the plots, the lower bounds of the moments display extremely rich structures, including sharp turning points at smaller values of~$\Delta_{\phi}$. These patterns are particularly intriguing: instead of carving out theories by maximizing a gap, here we are \emph{minimizing} the moments, essentially probing the opposite side of the bootstrap landscape. Understanding these minima requires taking into account the dynamics of extremely light operators, leading us into a region of the bootstrap that, to our knowledge, remains largely unexplored. Section~\ref{sec:new-kinks} is devoted to analyzing these new bootstrap structures through their extremal moments and spectra.

\section{The Ising moments}
\label{sec:ising}

The 3d Ising model has played a central role in studies of the numerical conformal bootstrap. In the work of~\cite{El-Showk:2012cjh}, the critical Ising model was identified as a sharp kink on the boundary of the allowed theory space, corresponding to gap maximization of the $\epsilon$ operator in the $\sigma \times \sigma$ OPE. In this section, we revisit the bootstrap for the 3d Ising model in the language of moment variables. These quantities capture coarse-grained features of the CFT spectrum and provide an alternative geometric perspective of theory space. We find that, in the space of low-lying moments, the upper bounds on these observables also develop a sharp kink precisely at the 3d Ising value. We present the numerical results, draw connections to known extremal properties of the Ising spectrum, and finally compare our best determination of the moments with those computed from previous Ising data.

\subsection{Correlator bounds}

\begin{figure}[tbp]
	\centering
	\includegraphics[width=0.9\textwidth]{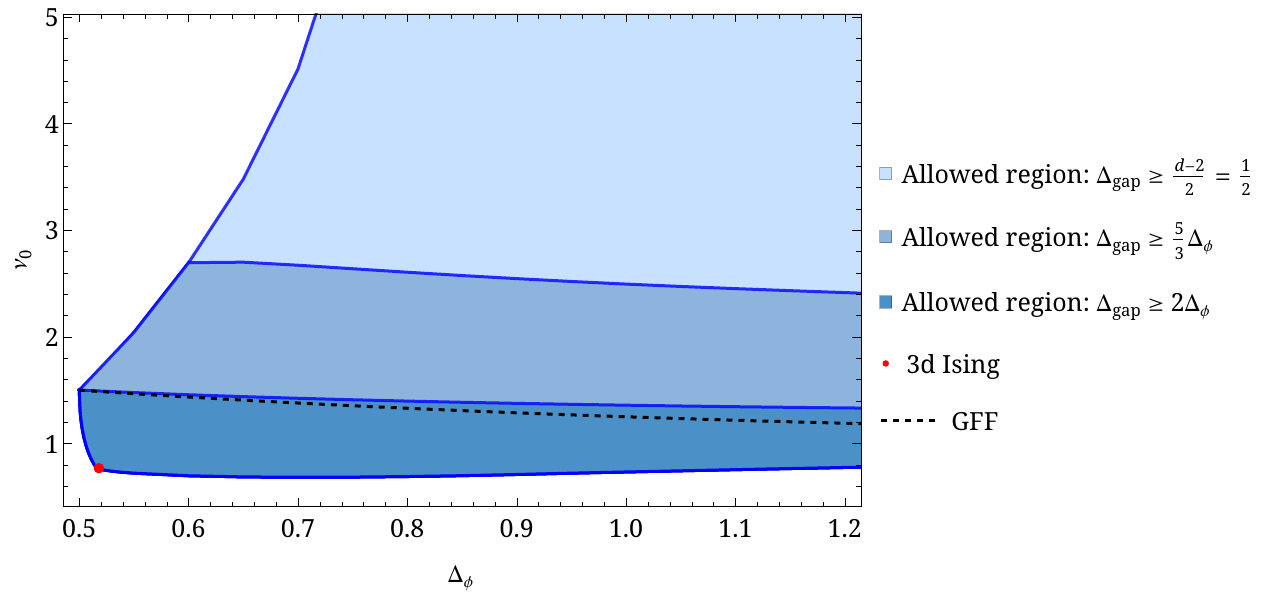}
	\caption{Reproduction of the correlator bounds in $d=3$ as a function of the external scaling dimension $\Delta_\phi$, following~\cite{Paulos:2021jxx}.	The sharp leftmost edge, where all bounds converge, corresponds to the free theory, while the red dot on the lower bound marks the 3d Ising correlator, computed from the conformal data of~\cite{Simmons-Duffin:2016wlq}. The bounds were obtained at $\Lambda = 11$.}
	\label{fig:correlator-bound}
\end{figure}

Correlator bounds in $d=3$ were previously examined in~\cite{Paulos:2021jxx}, which derived two-sided bounds and analyzed their dependence on the gap assumption. In particular, they identified a sharp kink in the lower bound of the correlator, which corresponds to the 3d Ising model, and related correlator extremization to the gap maximization. In our framework, the correlator is exactly the zeroth unnormalized moment, and we reproduce their bounds as shown in figure~\ref{fig:correlator-bound}. In the heavy correlator limit, the extremal solutions are saturated by the generalized free field correlators.

\subsection{Bounds on the $\Delta$-moments}

\begin{figure}[tbp]
	\centering
	\includegraphics[width=0.8\textwidth]{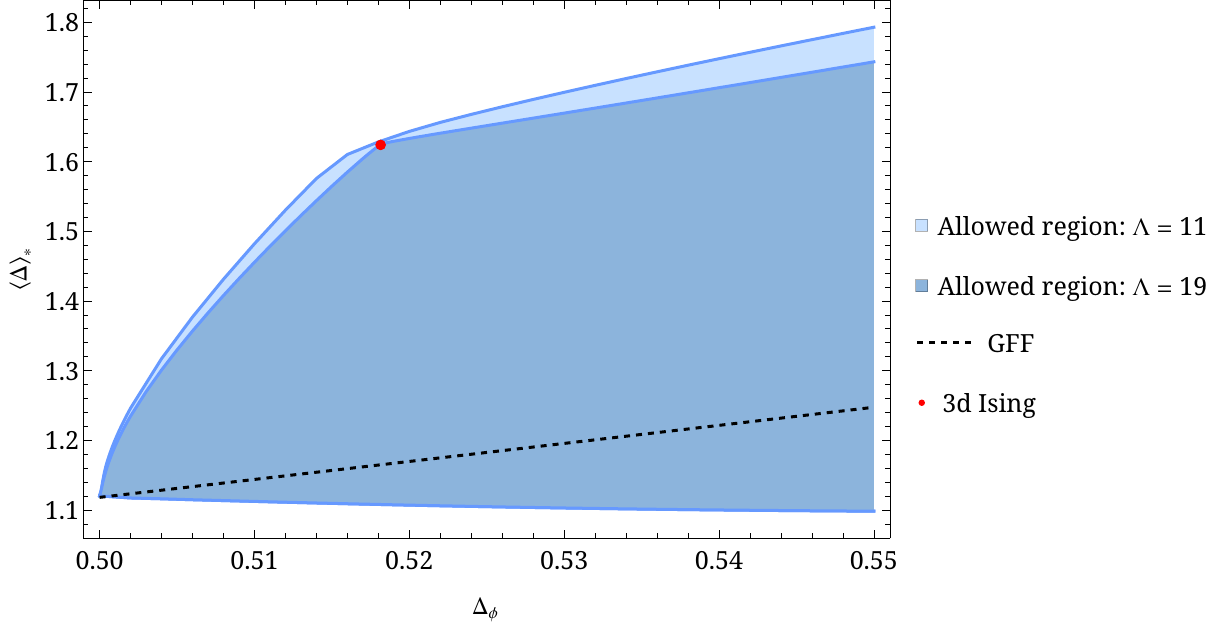}
	\caption{Upper and lower bounds on the first normalized moment $\langle \Delta \rangle_* = \nu_1 / (\nu_0 - 1)$ at $\Lambda = 11, 19$. The Ising model (red dot) sits at the kink.}
	\label{fig:Ising_1}
\end{figure}

It is not only the zeroth moment that places known theories on the boundary of the allowed region. Bounds on the higher moments of scaling dimension reveal additional geometric features, including the sharp corners associated with the Ising model. Here, we consider the first normalized $\Delta$-moment without the inclusion of the identity operator, $\langle \Delta \rangle_*$, with two-sided bounds shown in figure~\ref{fig:Ising_1}. A sharp kink is located at $\Delta_\phi \approx 0.518$. Similarly, the second moment $\langle \Delta^2 \rangle_*$ exhibits a kink at the same location.

\begin{figure}[tbp]
	\centering
	\includegraphics[width=0.82\textwidth]{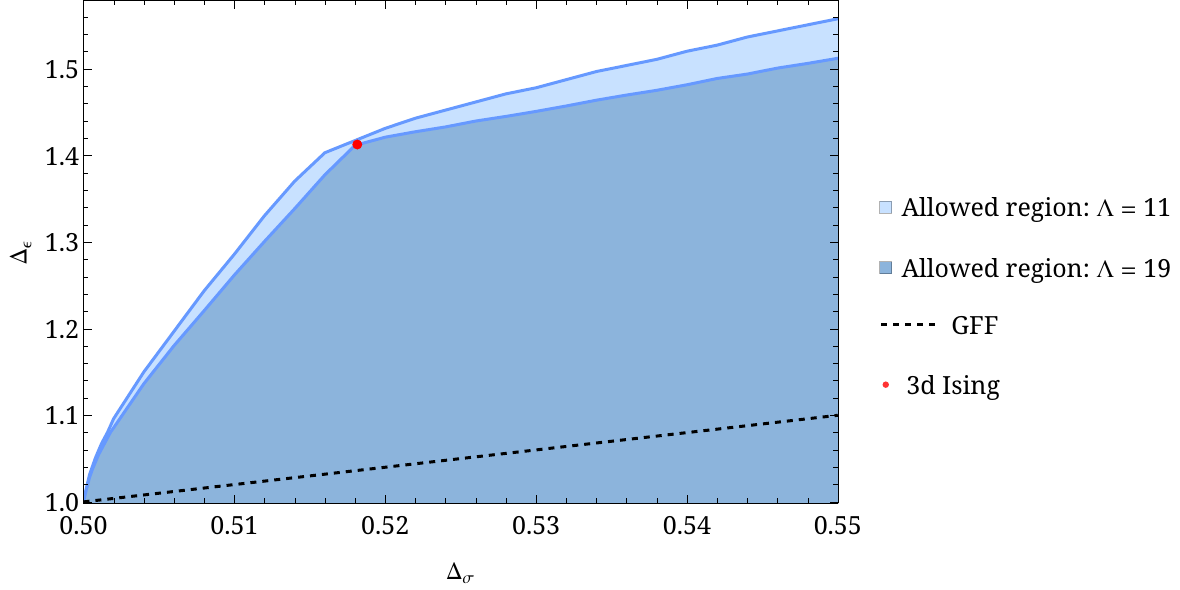}
	\caption{Gap maximization at $\Lambda = 11, 19$. The Ising model (red dot) sits at the kink.}
	\label{fig:Ising_max_gap}
\end{figure}

The emergence of the kink in the leading $\Delta$-moments is intuitive. For light correlators, the extremal solution is dominated by the gap state, which in this case corresponds to the $\epsilon$ operator in the $\sigma \times \sigma$ OPE. Since the 3d Ising model maximizes the gap and sits precisely at the kink of the gap-maximization plot, the $\Delta$-moment bounds naturally reflect this property. For reference, figure~\ref{fig:Ising_max_gap} shows the gap-maximization results at the same derivative order, allowing for a direct comparison between the two approaches. This pattern persists for higher moments, although the kink becomes less sharp, as heavier operators contribute more significantly to the weighted averages. Plots for moments across different dimensions can be found in appendix~\ref{appendix:Ising_across_dims}.

\subsection{Bounds on the $\ell$-moments}
\begin{figure}[tbp]
	\centering
	\includegraphics[width=0.85\textwidth]{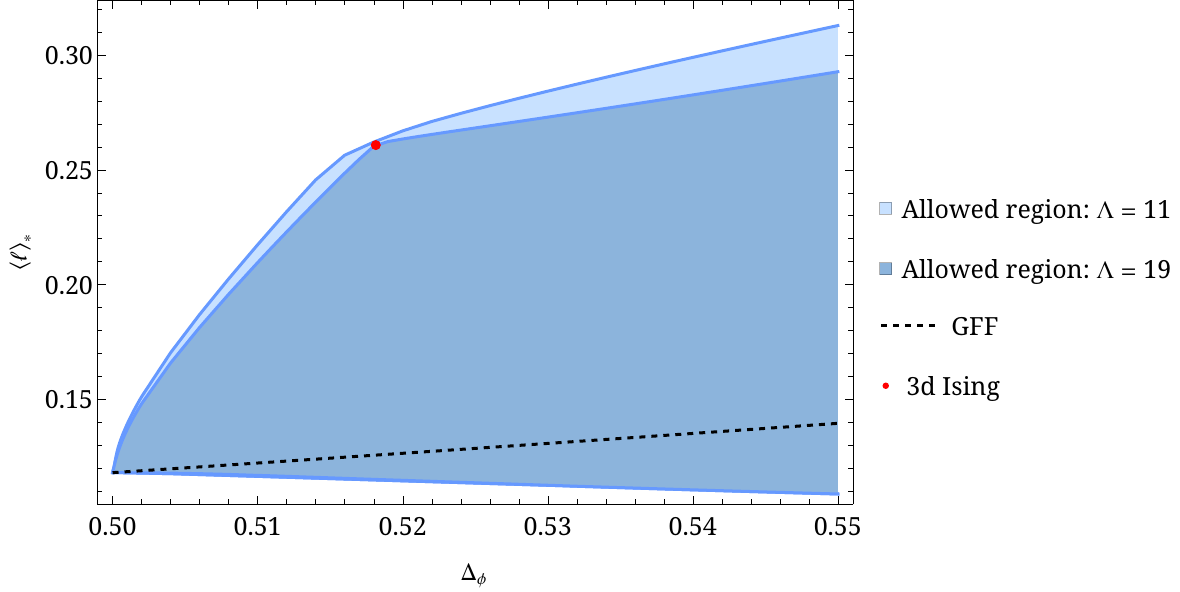}
	\caption{Two-sided bounds on the first normalized moment of spin, $\langle \ell \rangle = \nu_{0,1}/(\nu_{0,0}-1)$, excluding the identity operator contribution. The upper bound features a sharp kink precisely at the 3d Ising point.}
	\label{fig:Ising_01}
\end{figure}

In complete analogy with the previous analysis, one can also bound the moment variables associated with spin. Figure~\ref{fig:Ising_01} shows two-sided bounds on the first normalized $\ell$-moment, $\langle \ell \rangle_* = \nu_{0,1} / (\nu_{0,0}-1)$. Intriguingly, much like the gap-maximization and $\Delta$-moment bounds, the upper bound develops a sharp kink precisely at the 3d Ising point. The strictly positive lower bound also proves the necessary existence of spinning operators in the OPE.

\begin{figure}[tbp]
	\centering
	\includegraphics[width=0.85\textwidth]{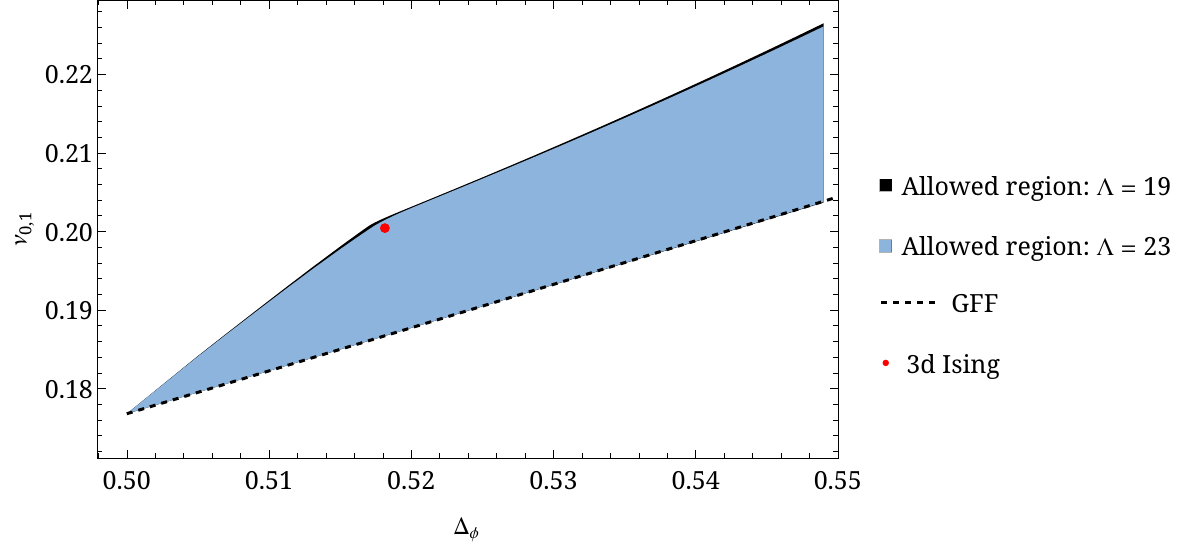}
	\caption{Allowed region for the first unnormalized moment of spin, $\nu_{0,1}$. Again, a kink can be seen close to the 3d Ising point. The lower bound is saturated by the corresponding moment of the generalized free field theory correlator $\langle\phi\phi\phi\phi\rangle$.}
	\label{fig:Ising_01_unnormalized}
\end{figure}

The appearance of the kink in the $\ell$-moment upper bound is interesting, as it has no direct analogue in the conventional gap-maximization analysis. To gain further insight, let us instead examine the unnormalized moment $\nu_{0,1}$, whose two-sided bounds are shown in figure~\ref{fig:Ising_01_unnormalized}. The fact that the Ising model maximizes the spin moment is intuitive: since $\nu_{0,1}$ receives contributions exclusively from spinning operators, the stress tensor provides the dominant contribution. The moment variable therefore attains its maximum when the stress tensor couples most strongly to the system, i.e., when the central charge reaches its minimal value. The formation of the kink, however, reflects a characteristic phenomenon of operator decoupling, closely tied to the underlying equation of motion in the Wilson–Fisher theory. As observed in \cite{El-Showk2014-ce}, the scaling dimensions of the spin-2 operators in the extremal spectra undergo a sudden rearrangement upon crossing the Ising point. The moment-maximizing spectra exhibit the same behavior, giving rise to the kink observed in the bounds.


\subsection{Precise moments under a gap assumption}
Finally, we present our most precise numerical bounds on the unnormalized moments obtained under a fixed external scaling dimension and scalar gap,
\begin{equation}
	\Delta_{\sigma} = 0.5181488, \qquad \Delta_{\epsilon} = 1.412625,
\end{equation}
in 3d, corresponding to the current best estimate of the scaling dimensions of the relevant operators in the 3d Ising CFT~\cite{Chang:2024whx}. This aggressive gap assumption pushes the bootstrap solution very close to the boundary of the allowed theory space, so that the resulting moments are expected to be tightly constrained around their true values.

Table~\ref{tab:ising_moments} summarizes the two-sided bounds on the lowest unnormalized moments~$\nu_{m,n}$. As can be seen, these bounds are indeed very tight, in some cases even excluding the values computed from the stable operators in \cite{Simmons-Duffin:2016wlq}, which are highlighted in red. This is not unexpected, since the list of \emph{stable operators} in \cite{Simmons-Duffin:2016wlq} is not expected to include all operators contributing to the moments. In particular, higher-twist trajectories and unstable operators may give non-negligible contributions. Quite remarkably, our numerical moment bounds, obtained at a significantly lower derivative order, are already sensitive enough to detect the effects of these missing operators.

As we have seen, the moment bounds presented here resemble the familiar results from gap maximization and $c$-minimization, yet they encapsulate both within a single, unified framework. However, they differ in two crucial aspects.
\begin{enumerate}
	\item The moment variables are defined as weighted averages over the entire operator spectrum. The resulting bounds therefore do not merely constrain a single operator, such as the gap state, but instead impose rigorous conditions on the collective behavior of all operators above the gap. The moment bootstrap carves out the theory space by characterizing the coarse-grained behavior of an infinite number of operators, and remarkably, this can be achieved numerically, providing rigorous statements about the full spectrum even at finite derivative order. As we have seen in section~\ref{sec:heavy correlator}, even in the heavy correlator regime, where the spectrum becomes dense, the numerical moment bounds remain well behaved and continue to yield meaningful constraints.
	\item As is evident from the plots, the moment variables are bounded not only from above, but also from below. As the external scaling dimension increases, the lower bounds deviate progressively from the generalized free field solutions and develop rich, highly nontrivial structures, including previously unobserved kinks. Section~\ref{sec:new-kinks} is devoted to a detailed analysis of these new structures and their physical interpretation.
\end{enumerate}

\begin{table}[tpb]
	\centering
	\begin{tabular}{c c c}
		\hline
		Moment               & Bounds        & Computed from~\cite{Simmons-Duffin:2016wlq} \\
		\hline
		$\nu_{0,0}$          & 0.768530(25)  & 0.7685483(69)                               \\
		$\nu_{1,0}$          & 1.247745(15)  & 1.247736(21)                                \\
		$\nu_{0,1}$          & 0.2004013(17) & \textcolor{red}{0.200381(14)}               \\
		$\nu_{2,0}$          & 2.2725170(20) & \textcolor{red}{2.272380(91)}               \\
		$\nu_{1,1}$          & 0.6255126(41) & \textcolor{red}{0.625408(68)}               \\
		$\nu_{0,2}$          & 0.4246333(36) & \textcolor{red}{0.424550(54)}               \\
		$\nu^{(\ell=0)}_{0}$ & 0.671220(26)  & 0.6712437(60)                               \\
		$\nu^{(\ell=0)}_{1}$ & 0.949868(19)  & 0.949883(12)                                \\
		$\nu^{(\ell=0)}_{2}$ & 1.3482213(37) & 1.3482174(31)                               \\
		$\nu^{(\ell=2)}_{0}$ & 0.0945027(28) & 0.09450447(26)                              \\
		$\nu^{(\ell=2)}_{1}$ & 0.2836123(20) & 0.28361131(80)                              \\
		$\nu^{(\ell=2)}_{2}$ & 0.8513774(44) & 0.8513735(25)                               \\
		\hline
	\end{tabular}
	\caption{
		Two-sided bounds on the leading unnormalized moments~$\nu_{m,n}$ and fixed-spin moments $\nu^{(\ell)}_{k}$, obtained with truncation parameters $\Lambda = 27$, $\ell_{\text{max}} = 50$, $r_{\text{max}} = 60$, and $\kappa = 20$. Moments whose bootstrap bounds do not overlap with the values reconstructed from the ``stable operators'' in the spectrum of~\cite{Simmons-Duffin:2016wlq} are highlighted in red.
	}
	\label{tab:ising_moments}
\end{table}

\section{A tour of the moment landscape}
\label{sec:new-kinks}

In the previous sections, we have shown that moment variables are an efficient way of probing the space of heavy correlators numerically and can also provide a useful characterization of the $\langle\sigma\sigma\sigma\sigma\rangle$ correlator in the critical Ising model. While these results are already encouraging, the true strength of the moment bootstrap lies in its ability to uncover previously unexplored structures in the space of solutions to the crossing equations.

In this section, we guide the reader through the rich set of geometric features that characterize our moment bounds. Such structures arise only along the \emph{lower bounds} of the moments for light correlators, as can be seen in figure~\ref{fig:heavy moment-bounds}, long before the bounds converge to the smooth power laws predicted analytically in the heavy correlator limit. This places us in a regime where the pole structure of the conformal blocks becomes important, and the extremal spectra reorganize in highly nontrivial ways.

As we will demonstrate, the geometric structures in the moment bounds are closely related to several distinct features of the extremal spectrum, including
\begin{itemize}
	\item operator decoupling of low-lying operators,
	\item extremization of scaling dimensions,
	\item saturation of unitarity bounds, and
	\item evidence for an extremal spectrum that simultaneously extremizes multiple low-lying moment variables.
\end{itemize}
In the following sections, we take a guided tour of the moment landscape revealed by the numerical bootstrap, focusing on the range $2 < d < 6$ and on the relationship between these structures and the spectral properties listed above.

\subsection{Overview of geometric features}

\begin{figure}[tbp]
	\centering
	\includegraphics[width=\textwidth]{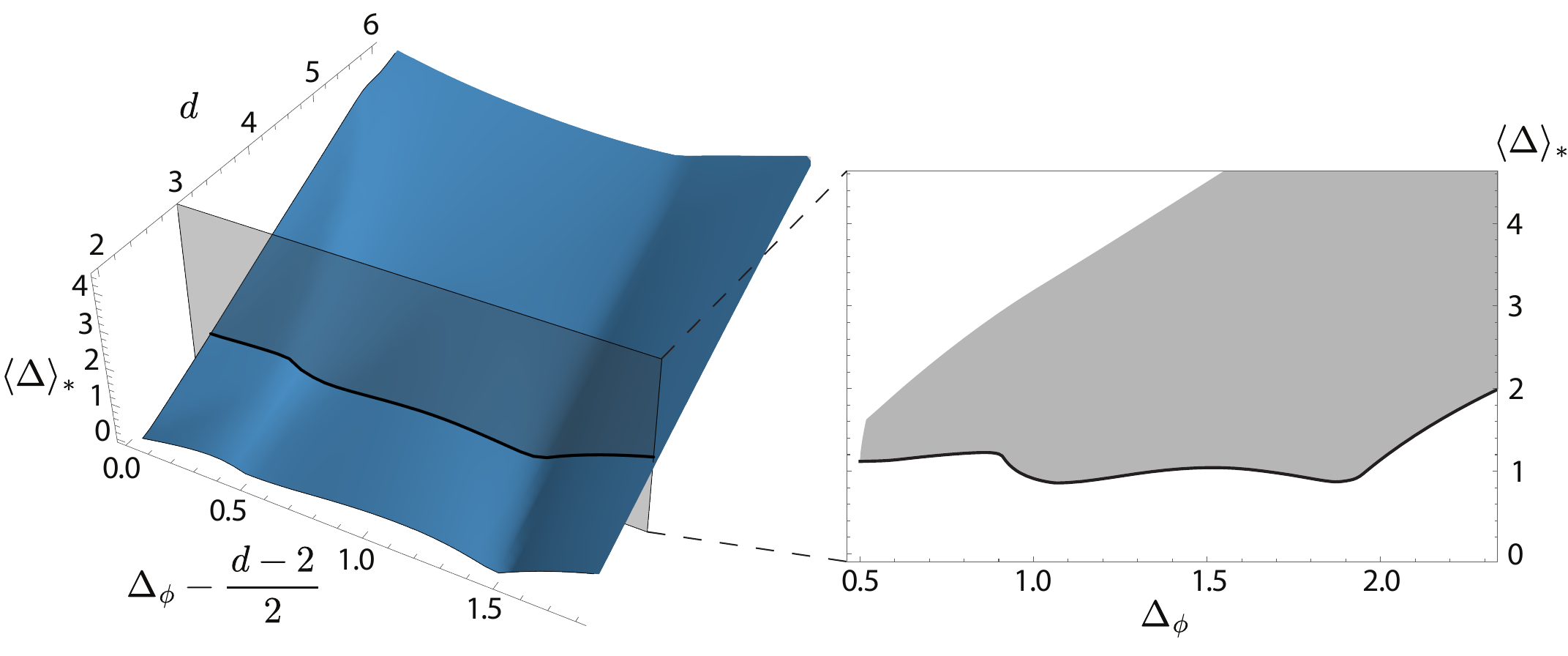}
	\caption{The lower bound on the normalized first moment $\langle \Delta \rangle_*$ as a function of $(d, \Delta_\phi - \frac{d-2}{2})$, with a cutaway of the allowed region cross-section at $d =3$, computed at $\Lambda = 23$. Two continuous families of kinks are clearly visible on the surface, corresponding to the sharp turning features seen in the contour plot, figure~\ref{fig:landscape_3dcontourplot}. The first family forms a smooth curve from $(d,\Delta_\phi) = (2^+, \tfrac{1}{2})$ to $(d,\Delta_\phi) = (6,2)$, where it merges into the free scalar theory. The second family originates from $(d,\Delta_\phi) = (2^+, \tfrac{3}{2})$ and persists toward higher dimensions. The features are not present in exactly $d=2$, which we discuss in section~\ref{sec:fake_primary}.
	}
	\label{fig:landscape_3dplot}
\end{figure}

\begin{figure}[tbp]
	\centering
	\includegraphics[width=0.7\textwidth]{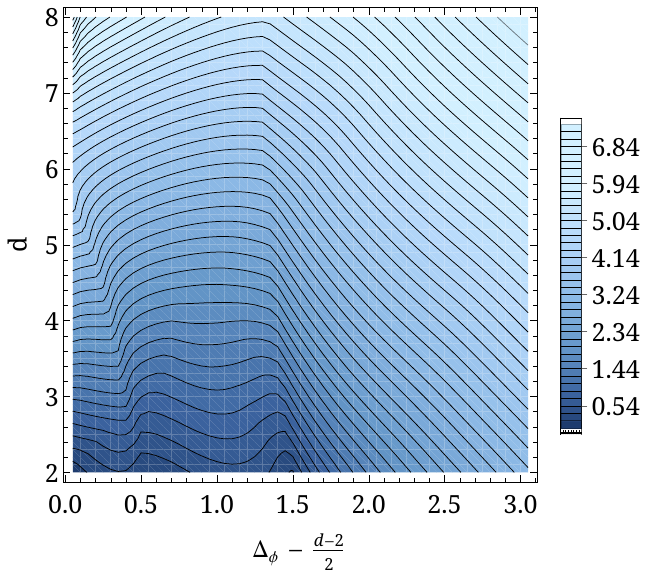}
	\caption{Contour plot of the same moment landscape shown in figure~\ref{fig:landscape_3dplot}. The first family of kinks (the ``cliff'') can be traced continuously from its endpoint at $d \rightarrow 2^+$ to the free scalar theory at $d=6$.}
	\label{fig:landscape_3dcontourplot}
\end{figure}

By examining the lower bounds on $\langle \Delta \rangle_*$, i.e., the first normalized moment without the inclusion of the identity, across a wide range of spatial dimensions, a global moment landscape emerges: the lower bound traces out a non-trivial surface in the $(d, \Delta_\phi)$ plane (figures~\ref{fig:landscape_3dplot} and~\ref{fig:landscape_3dcontourplot}), and we identify two families of kinks. As we will see shortly, they are closely related to discontinuous or extremizing features of the extremal spectra, in particular to decoupling events involving low-lying operators.

Across all dimensions $2 < d < 6$, the lower bound at fixed $d$ follows the same qualitative progression of geometric landmarks as $\Delta_\phi$ increases:
\begin{equation}
	\label{eq:universal-sequence}
	\begin{aligned}
		 & \text{cliff}
		\;\longrightarrow\;
		\underbrace{\text{first valley}
			\;\longrightarrow\;
			\text{hill}
			\;\longrightarrow\;
			\text{second valley}}_{\text{merge into a single ``crater'' in $d\gtrsim 4$}}
		\;\longrightarrow\;
		\text{asymptotically linear ramp}.
	\end{aligned}
\end{equation}

The two families of kinks remain robust throughout the entire range $2<d<6$, as can be clearly traced in figures~\ref{fig:landscape_3dplot} and~\ref{fig:landscape_3dcontourplot}, occurring near the ``cliff'' and again near the onset of the asymptotically linear ramp.
The intermediate features, by contrast, are more malleable: in $d \lesssim 4$, they appear as two well-separated local minima with a peak between them, while in $d \gtrsim 4$ they merge into a broad, single ``crater'' that eliminates the W-shaped lower bound, with the two families tracing roughly the two sides of the crater.

As $d\to 2^+$, the lower bounds are \emph{discontinuous}: in exactly $d=2$, the lower bound shows no kinks and closely tracks the linear trajectory $\sqrt{2}\,\Delta_\phi$ (figure~\ref{fig:summary_2d}). We will return to this phenomenon in section~\ref{sec:fake_primary}.

\begin{figure}[h!]
	\centering
	\includegraphics[width=\textwidth]{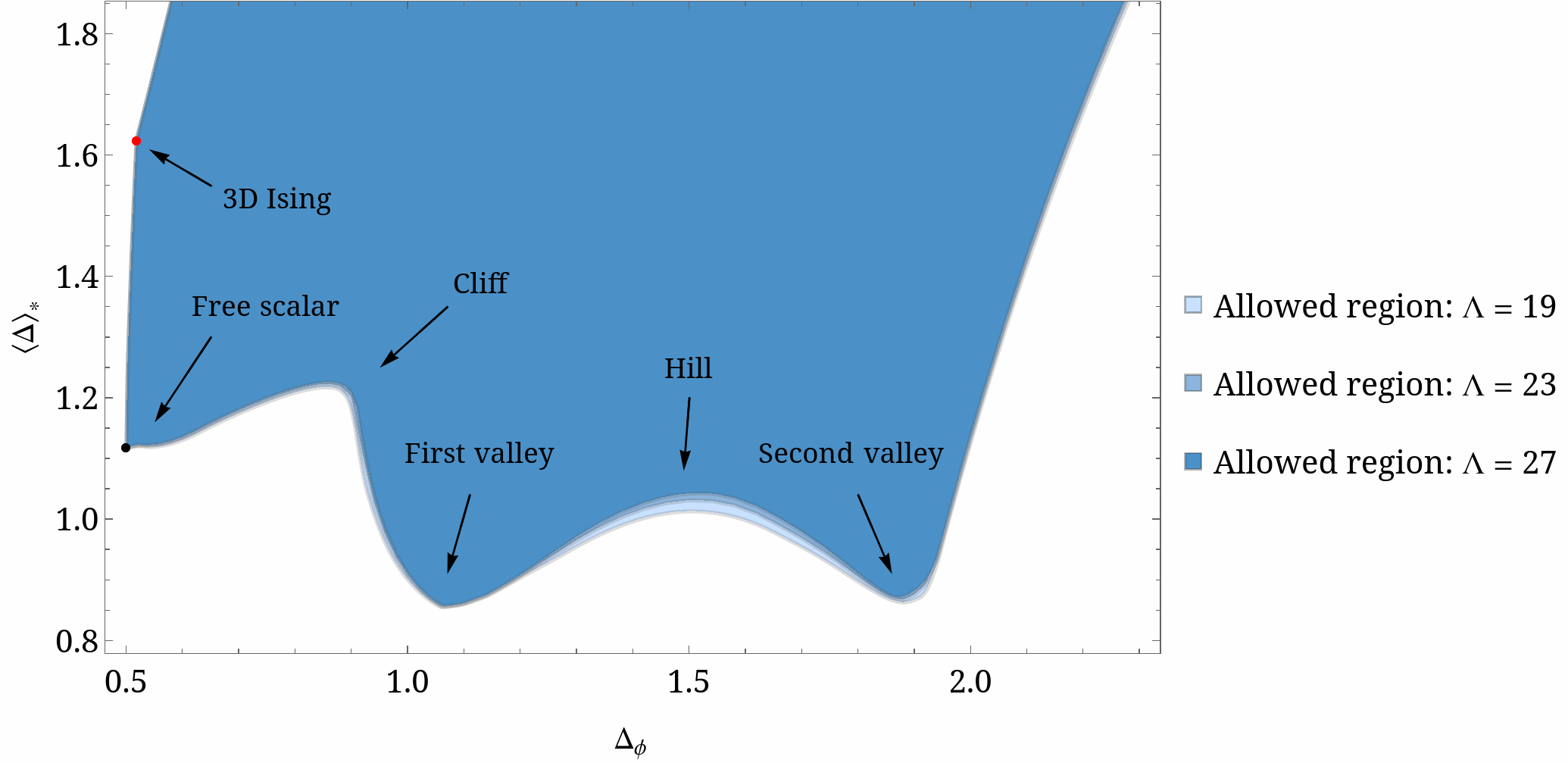}
	\caption{
		Lower bound on the normalized first moment $\langle \Delta \rangle_*$ in $d=3$. The W-shaped bound consists of a sharp ``cliff'' followed by two ``valleys'' separated by an intermediate ``hill.'' The qualitative structure and locations of these features remain stable under increases in the derivative order~$\Lambda$ and the spin truncation~$\ell_{\max}$.
	}
	\label{fig:kinks_3d}
\end{figure}

To discuss the spectral origin of the bootstrap landscape, it is convenient to zoom in on a representative slice. Figure~\ref{fig:kinks_3d} displays the lower bound on $\langle \Delta \rangle_*$ in $d=3$, where the W-shaped profile is stable under increases in the derivative order, $\Lambda$, and the spin truncation, $\ell_{\max}$, indicating that the features are genuine consequences of the bootstrap equations rather than numerical artifacts. Unless stated otherwise, the moment bounds and extremal spectra in the following subsections are obtained in $d=3$; deformations of these structures with the spatial dimension will be discussed explicitly where relevant.

\subsection{Spectral reorganization before the ``cliff''}

Although the moment bound presented earlier exhibits dramatic nonlinear behavior only after the ``cliff,'' significant spectral reorganization driven by the lightest scalar has already occurred beforehand.

\begin{figure}[tpb]
	\centering
	\includegraphics[width=\textwidth]{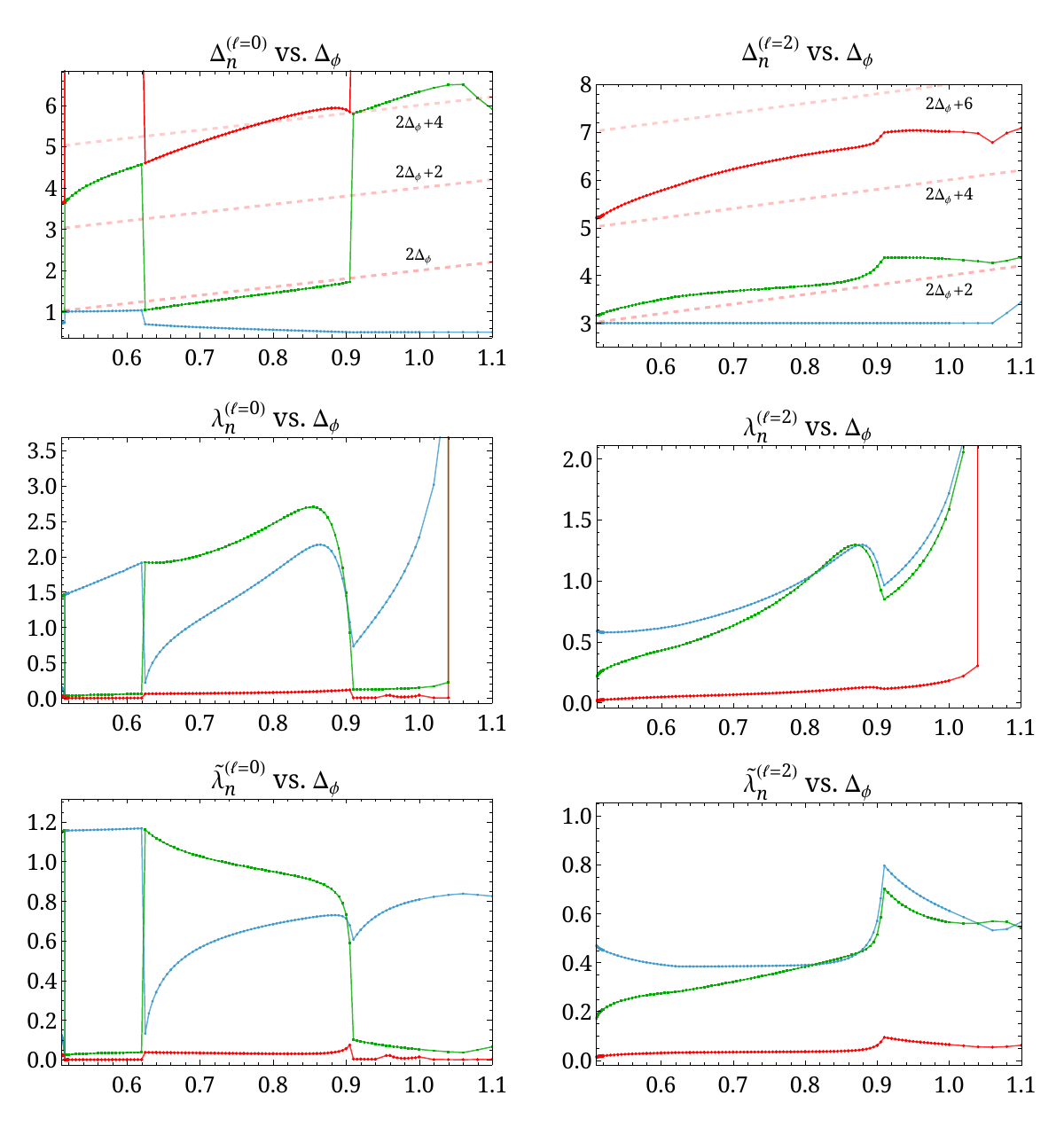}
	\caption{
		Extremal scaling dimensions and (normalized) OPE coefficients (see section~\ref{numerical_bootstrap_setup} for definition) of the leading scalar and spin-2 operators in $d=3$ as functions of $\Delta_\phi$, computed along the $\Lambda = 23$ lower bound of the normalized first moment. The blue, green, and red curves correspond to the lightest ($n=1$), second-lightest ($n=2$), and third-lightest ($n=3$) operators, respectively.
		At $\Delta_\phi \approx 0.514$, a light scalar $\varphi$ decouples.
		At $\Delta_\phi \approx 0.62$, $\varphi$ reappears and its scaling dimension subsequently drifts toward the unitarity bound.
		Near $\Delta_\phi \approx 0.91$, corresponding to the ``cliff'', the second-lightest scalar decouples.
		At $\Delta_\phi \approx 1.05$, the first ``valley'' is reached, where the reconstructed correlator diverges, and so do the unnormalized OPE coefficients. The normalized OPE coefficients, however, remain perfectly finite. 
	}
	\label{fig:first_op_decoup}
\end{figure}

The extremal spectra in a representative slice, $d=3$, are shown in figure~\ref{fig:first_op_decoup}, where we plot the scaling dimensions and (normalized) OPE coefficients of the lightest scalar and spin-2 operators as functions of the external scaling dimension $\Delta_{\phi}$. As we move away from the free scalar point, a light scalar operator begins to contribute to the OPE with a very small coefficient. Let us denote this operator by $\varphi$. It remains in the spectrum until $\Delta_\phi \approx 0.514$, where its OPE coefficient drops to zero and the operator decouples.

Beyond this point, the gap state is replaced by a scalar whose scaling dimension closely follows the GFF trajectory $\Delta \approx 2\Delta_\phi$, which we schematically denote by $\phi^2$. At $\Delta_\phi \approx 0.62$, interestingly, the light scalar $\varphi$ reappears in the spectrum. From there onward, its scaling dimension gradually approaches the unitarity bound, and its contribution to the OPE eventually becomes dominant.

\begin{figure}[tpb]
	\centering
	\includegraphics[width=0.7\linewidth]{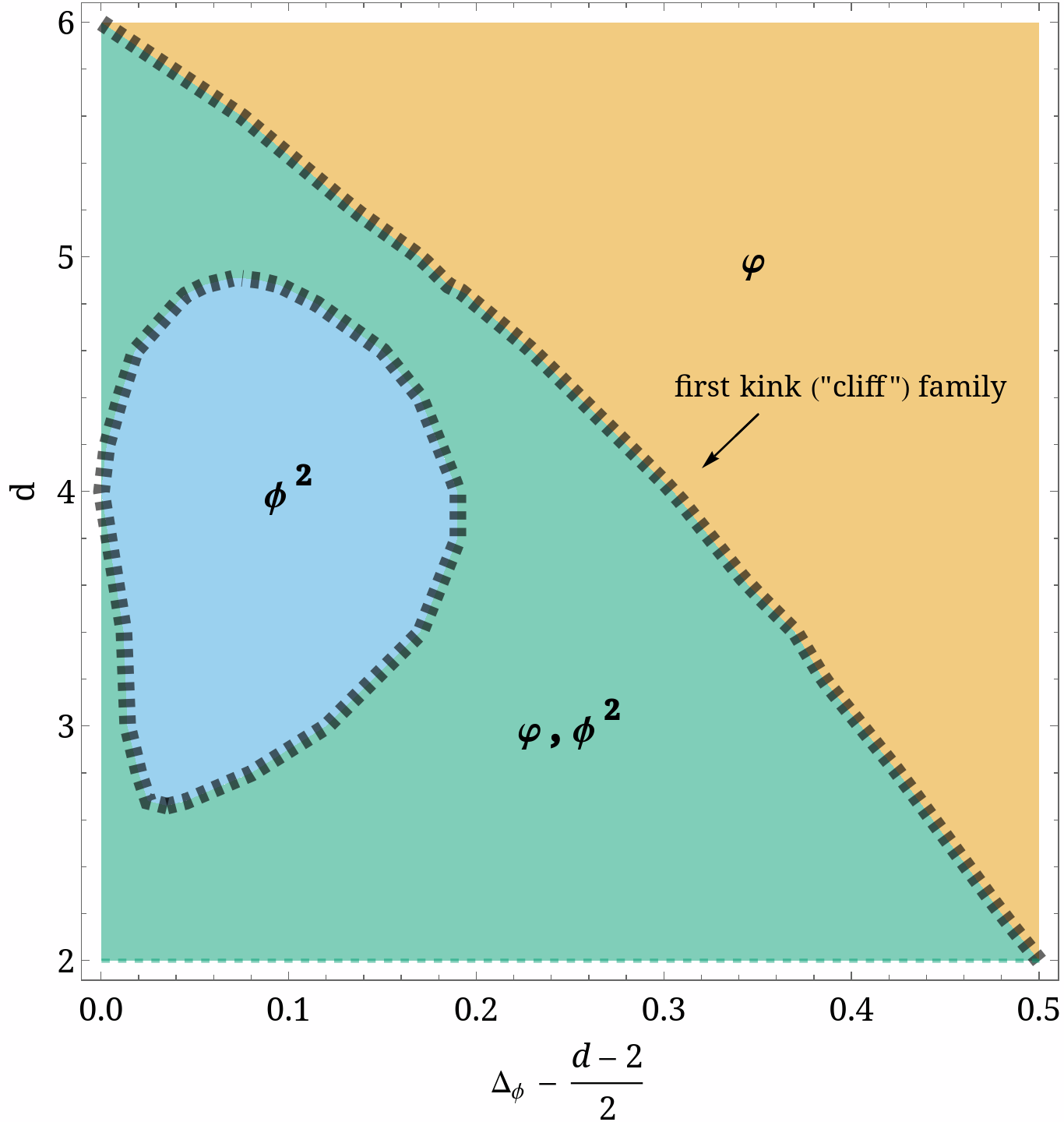}
	\caption{
		A map of the ``CFT park'' illustrating the landscape of the lightest scalar operator spectrum prior to the first kink family (the ``cliff''). The colors indicate qualitatively distinct regions, with $\varphi$ representing a light scalar whose scaling dimension eventually drifts toward the unitarity bound, and $\phi^2$ an operator whose scaling dimension closely follows the $2\Delta_{\phi}$ trajectory. The intermediate ``pond'' region (in blue) is characterized by the absence of the light scalar $\varphi$. The dashed curves mark the corresponding boundaries at which operator decoupling occurs. We stress that the naming of these operators is descriptive, as no underlying perturbative explanation has been discovered.
	}
	\label{fig:cft_park_map}
\end{figure}

\paragraph{Deformation across dimensions} The decoupling phenomenon described above can be seen only within a finite window of spatial dimensions, $2.6 \lesssim d \lesssim 4.9$. We summarize the global structure in figure~\ref{fig:cft_park_map}, where regions with different colors correspond to extremal spectra with qualitatively different scalar operator content.

Within this window, there exists a special region that we refer to as the ``pond'', where the scalar gap state is the $\phi^2$ operator instead of the light scalar $\varphi$. Within our numerical resolution, the ``pond'' appears to touch the free scalar theory at $d = 4$, and is bounded within the interval $2.6 \lesssim d \lesssim 4.9$.

For $d \geq 5$, we do not observe any operator decoupling until the ``cliff.'' Instead, the low-lying conformal data exhibit qualitatively different behavior, producing another sharp corner in the moment bound, as shown in figure~\ref{fig:anomdim_d6}. We will return to these structures in the next section.

\subsection{The cliff}

\paragraph{Operator decoupling and ``gap-minimization''}
At the ``cliff'', located at $\Delta_{\phi} \approx 0.908$ in $d=3$, the second-lightest scalar suddenly decouples from the extremal spectrum, as shown in figure~\ref{fig:first_op_decoup}. This decoupling is accompanied by a sharp drop in its OPE coefficient and thus a suppression of its contribution to the spectral density~\eqref{def_spec_dens}. Meanwhile, the lightest scalar has a scaling dimension very close to the unitarity bound, $\frac{d-2}{2}$, which makes the corresponding conformal block extremely large, contributing significantly to the spectral density. Together, these effects produce the sudden drop in the moment.

An interesting feature of the extremal spectra is that the scaling dimension of the decoupled operator lies extremely close to $2\Delta_{\phi}$, which is the value of the generalized free field double-trace operator $[\phi\phi]_{0,0}$ at the same external weight. Furthermore, as shown in figure~\ref{fig:lightest_scalar_dim}, the already small anomalous dimension of the lightest scalar reaches its local minimum at precisely the same location, essentially ``minimizing'' the gap, revealing solutions on the opposite side of the usual bootstrap landscape!

\paragraph{Deformation across dimensions}
Interestingly, the ``cliff'' discussed above belongs to a continuous family of kinks across spatial dimensions, and a notable event occurs at $d = 6$. Here, this family collides with the free scalar theory point at $\Delta_\phi = \frac{6-2}{2} = 2$, terminating the family entirely, as can be seen in figures~\ref{fig:landscape_3dplot}, \ref{fig:landscape_3dcontourplot}, and~\ref{fig:cft_park_map}.

\begin{figure}[tpb]
	\centering
	\makebox[\textwidth][c]{%
		\begin{subfigure}[t]{0.48\textwidth}
			\centering
			\includegraphics[height=0.85\textheight,keepaspectratio]{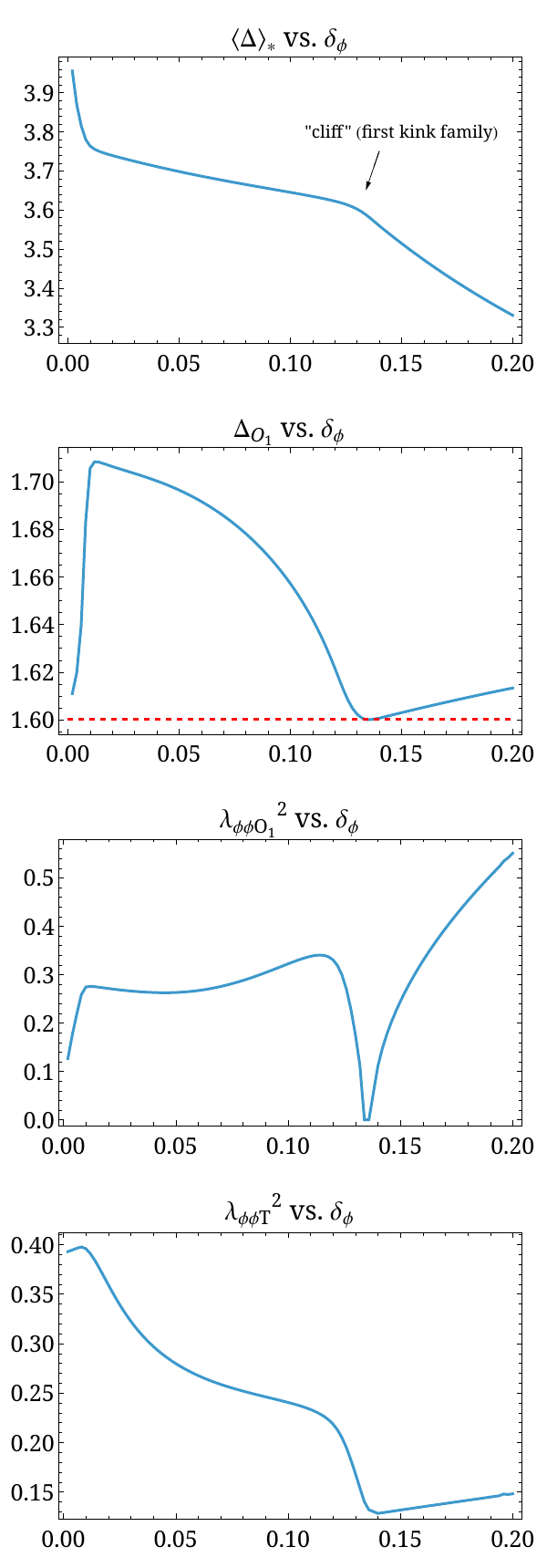}
			\caption{$d = 5.2$}
		\end{subfigure}
		\begin{subfigure}[t]{0.48\textwidth}
			\centering
			\includegraphics[height=0.85\textheight,keepaspectratio]{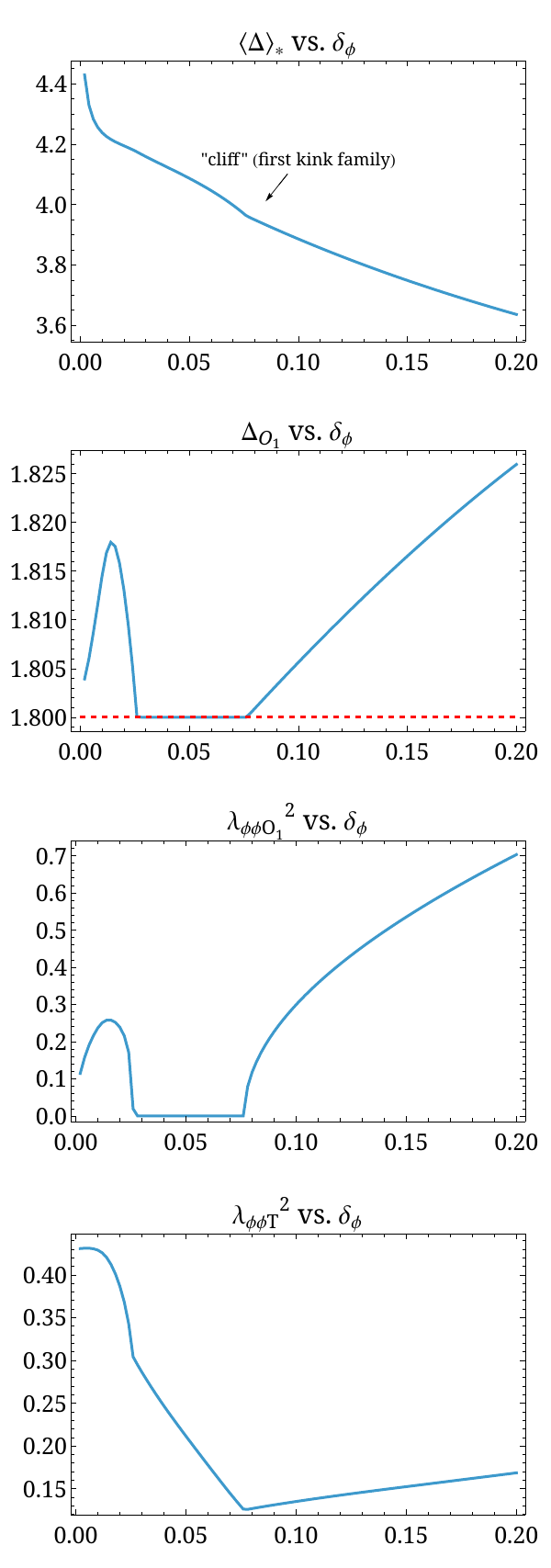}
			\caption{$d = 5.6$}
		\end{subfigure}
	}
	\caption{Moment lower bounds (obtained at $\Lambda = 23$) and corresponding extremal spectra in representative spatial dimensions near $d=6$, plotted as functions of $\delta_{\phi} \equiv \Delta_{\phi} - \frac{d-2}{2}$. The red dashed line denotes the unitarity bound. In addition to the ``cliff'' associated with an operator decoupling, we observe a preceding sharp corner. The latter coincides with local maxima of both the lightest-scalar scaling dimension and the central charge, while the cliff corresponds to local minima of the same quantities.}
	\label{fig:anomdim_d6}
\end{figure}

Near $d=6$, both the anomalous dimensions and the OPE coefficients of the lightest operators exhibit sharp, discontinuous behavior as we approach the free scalar theory, as shown in figure~\ref{fig:anomdim_d6}. Beyond $d\gtrsim 5$, we consistently see two distinguished corners along the lower bound. The first corner is not accompanied by an operator decoupling and coincides with a local maximum of the lightest-scalar anomalous dimension and of the central charge. By contrast, the second corner belongs to the ``cliff'' family, which continuously connects to the operator decoupling above at $d=3$; there we find a local minimum of the lightest-scalar anomalous dimension and of the central charge.

Finally, we comment on the operator-decoupling event in general dimensions, responsible for the ``cliff'' observed. As pointed out, the scaling dimension of the second-lightest scalar nearly sits at the generalized free field value $2 \Delta_{\phi}$ at the cliff, but it does not coincide with this value uniformly across dimensions. We observe a small positive anomalous dimension for $d < 3$ and a negative anomalous dimension for $d > 3$.

\subsection{First valley}

Shortly after the ``cliff'', the moment $\langle \Delta \rangle_*$ reaches its first local minimum. This occurs where the contribution of the lightest scalar to the spectral density is dominant, and where the scaling dimension of the second-lightest scalar is locally maximized. Notably, the anomalous dimension of the lightest scalar also reaches a local maximum at the same location. These behaviors are shown in figures~\ref{fig:first_op_decoup} and~\ref{fig:lightest_scalar_dim}.

In $d=3$, the first valley marks a special point where the moment bootstrap begins finding qualitatively different moment-minimizing solutions. Using the spectrum-extraction procedure described in section~\ref{numerical_bootstrap_setup}, we can explicitly reconstruct the correlation functions from the extremal spectra along the lower bound. We find that upon passing the first valley, the reconstructed correlation function exhibits behavior consistent with becoming unbounded above. This is reflected in the fact that when the primal-dual gap in \texttt{SDPB} is tightened, the reconstructed correlator continues to grow. The unboundedness of correlators persists until the second valley is reached, where the reconstructed correlation function returns to a finite value.

The possibility that the correlation function may be unbounded above once a certain threshold $\phi \gtrsim d-2$ is crossed has been noted previously in~\cite{Paulos:2021jxx}. This behavior admits a natural interpretation: beyond the threshold, the numerical bootstrap is able to construct crossing-symmetric and unitary solutions that contain no identity operator. Such solutions can be rescaled by an arbitrarily large positive factor while preserving crossing symmetry and unitarity, rendering the correlation function unbounded above.

Although at finite numerical precision, the reconstructed correlator is merely extremely large rather than strictly infinite, its magnitude increases as the duality gap is reduced. This means that the numerical bootstrap is probing solutions in which the identity operator is negligible. More precisely, the moment-minimizing solution is effectively obtained by the bootstrap through increasing the overall normalization of the non-identity contribution to the correlator while keeping the ratio $\nu_1/(\nu_0 - 1)$ minimized. In this limit, only the normalized spectral density $\tilde{p}_\ell(\Delta)$ or the normalized OPE coefficient $\tilde\lambda_{\phi\phi\mathcal{O}}$, introduced in section~\ref{numerical_bootstrap_setup}, remains meaningful, as it divides the conformal data by the correlation function (without the identity contribution). 

\paragraph{Deformation across dimensions}
So far, we have not observed any clear connection between the unbounded correlator and the geometric features of the lower bound. Moreover, for $d \geq 5$, the reconstructed correlation function remains finite throughout the region we examine. The fact that moment-minimizing solutions in higher dimensions do not appear to favor identity-suppressed directions is intriguing and merits further investigation.

The first valley is also related to the presence of the stress tensor, which serves as a diagnostic for locality. In $d \le 3$, the first valley marks the point where the stress tensor decouples from the extremal spectrum. In $d>3$, local theories emerge after the first valley and occupy the lower bounds more and more as $d$ increases. Near $d = 6$, the stress tensor is present across the entire lower bound until the second valley.

\subsection{The hill}

\begin{figure}[tbp]
	\centering
	\includegraphics[width=\textwidth]{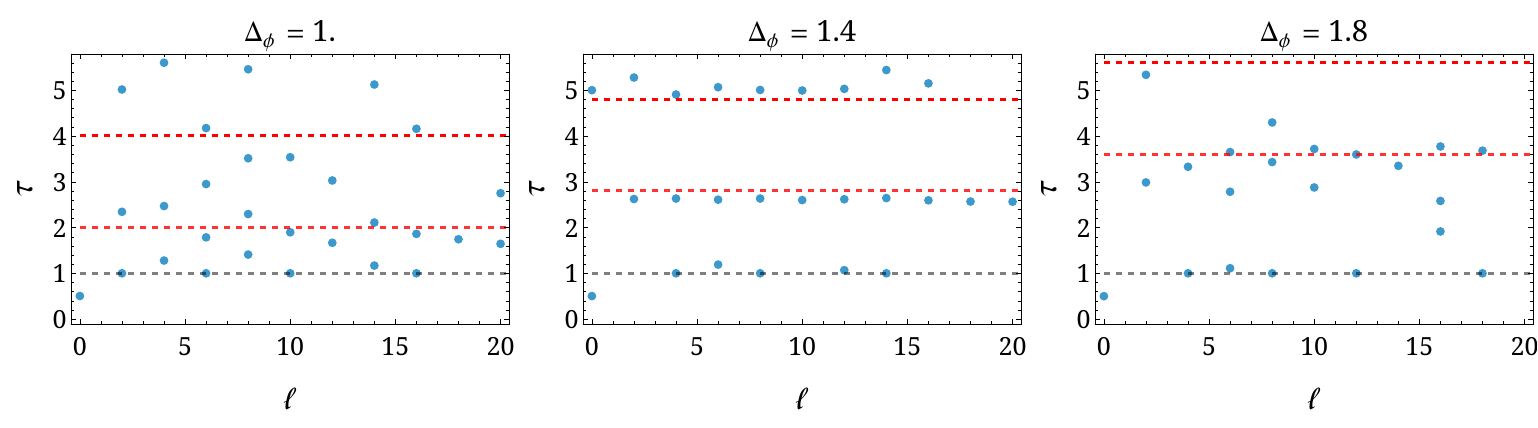}
	\caption{Formation and collapse of the constant-twist trajectories between the two valleys in $d = 3$, computed at $\Lambda = 23$. The red dashed lines indicate the leading GFF double-twist trajectories, $2\Delta_{\phi}$ and $2\Delta_{\phi}+2$, and the black dashed line the unitarity bound, $d-2=1$. On the ``hill’’ of the moment lower bound, higher-spin operators align along near-GFF double-twist trajectories: the leading family $[\phi\phi]_{0,\ell}$ exhibits small negative anomalous dimensions, while the next family $[\phi\phi]_{1,\ell}$ shows small positive anomalous dimensions. These trajectories are clearly visible in $2.8 \lesssim d \lesssim 4$.}
	\label{fig:hill_twist_traj}
\end{figure}

\paragraph{Emergence of constant twist trajectories} Between the two valleys, an interesting phenomenon appears in the extremal spectra: spinning operators begin to align along constant-twist trajectories that closely follow those of the GFF spectrum, up to small anomalous dimensions. In this regime, the leading double-twist family $[\phi\phi]_{0,\ell}$ typically acquires a small negative anomalous dimension, while the next family $[\phi\phi]_{1,\ell}$ shows a small positive anomalous dimension. These are illustrated in figure~\ref{fig:hill_twist_traj}.

\paragraph{Deformation across dimensions} The pattern above is consistently observed in the window $2.8 \lesssim d \lesssim 4$, which coincides with the range of spatial dimensions where the ``hill’’ in the moment landscape is clearly seen. Outside this window, both the hill and the associated twist trajectories become much less visible. A better understanding of this connection between the spectral and geometric properties of the moment bounds remains an open and intriguing direction for future work.

\subsection{Second valley}
In the second valley, where the reconstructed correlation functions return to finite values, several notable events occur in close succession:
\begin{enumerate}[label=(\alph*)]
	\item the lightest scalar saturates the unitarity bound, and
	\item the scaling dimension of the second-lightest scalar suddenly drops back to its previously decoupled value, $2\Delta_{\phi}$, closely followed by
	\item an extremal spectrum that minimizes all the leading moments simultaneously.
\end{enumerate}

\begin{figure}[tpb]
	\centering
	\includegraphics[width=\textwidth]{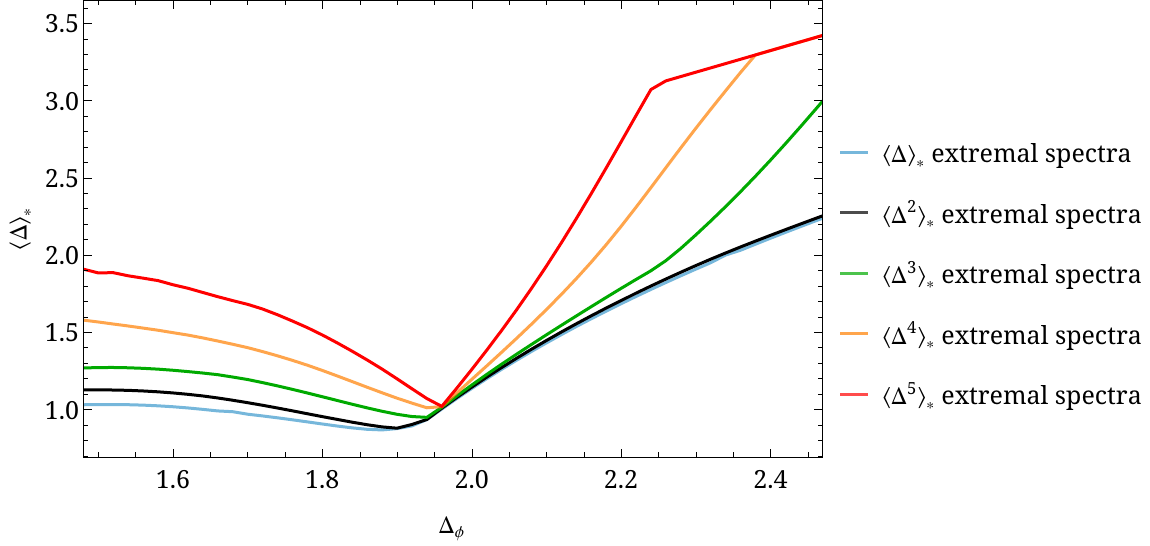}
	\caption{Shortly after the second valley, there is a special point where the moment-minimizing spectra coincide. This is illustrated by plotting the first moment evaluated using the extremal spectra data obtained by minimizing higher moments (here at $\Lambda =23$), shown in different colors. Around $\Delta_{\phi} \approx 1.96$, all extremal spectra share the same average scaling dimension. A similar analysis for the second moment is shown in figure~\ref{fig:simultaneous_minimization_second_moment}.}
	\label{fig:simultaneous_minimization_first_moment}
\end{figure}

\begin{figure}[tpb]
	\centering
	\includegraphics[width=\textwidth]{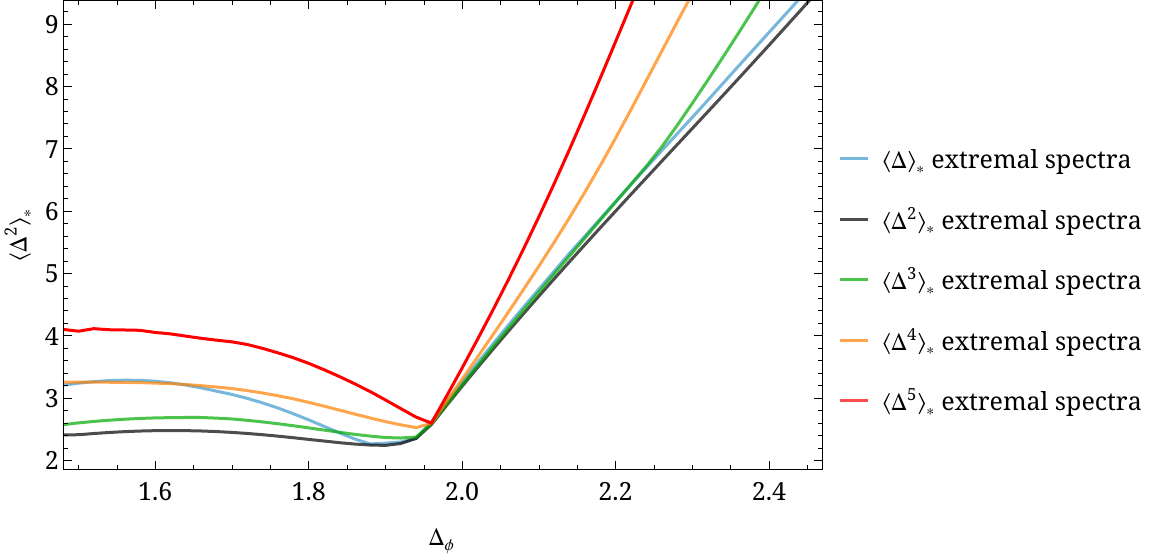}
	\caption{At the same location as in figure~\ref{fig:simultaneous_minimization_first_moment}, we plot the second moment, evaluated using the same extremal spectra obtained by minimizing different low-lying moments as in figure~\ref{fig:simultaneous_minimization_first_moment}. Near $\Delta_{\phi} \approx 1.96$, all extremal spectra yield to the same second moment. Similar behavior continues for higher moments.}
	\label{fig:simultaneous_minimization_second_moment}
\end{figure}

The behavior of the leading moments around point (c) is shown in figure~\ref{fig:simultaneous_minimization_first_moment} and figure~\ref{fig:simultaneous_minimization_second_moment}. Rather than comparing the extremal spectra directly, we use low-lying moments as efficient probes of spectral coincidence. Although the extremal spectra obtained by minimizing different moments appear scattered at the level of individual operators, their first and second moments collapse to the same value at point (c). Higher moments exhibit the same qualitative behavior, approximately coinciding at the same point as well. This strongly suggests the existence of a single extremal solution that simultaneously minimizes several low-lying moments. It would be interesting to study this point further in future work, either by increasing the number of derivatives or by applying the mixed-correlator bootstrap, which may help clarify the physical properties of this distinguished solution.


\begin{figure}[tpb]
	\centering
	\includegraphics[width=\textwidth]{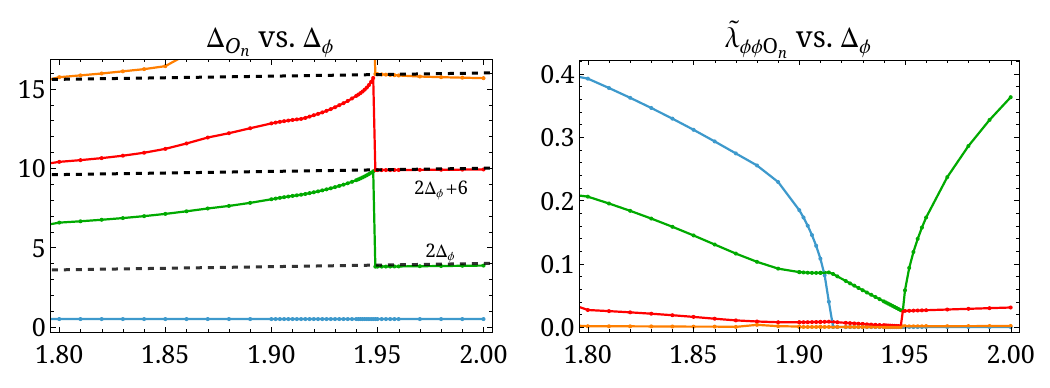}
	\caption{The scalar sector of the extremal spectra around the second valley, computed at $\Lambda = 27$. Two notable phenomena occur in close proximity: the lightest scalar approaches and saturates the unitarity bound, and the second-lightest scalar returns to $2\Delta_{\phi}$. The plots show the scaling dimensions (left) and the normalized OPE coefficients (right) of the four lightest scalar operators, labeled in blue ($n=1$), green ($n=2$), red ($n=3$), and orange ($n=4$).}
	\label{fig:second_kink_spec}
\end{figure}

\begin{figure}[tpb]
	\centering
	\includegraphics[width=0.8\textwidth]{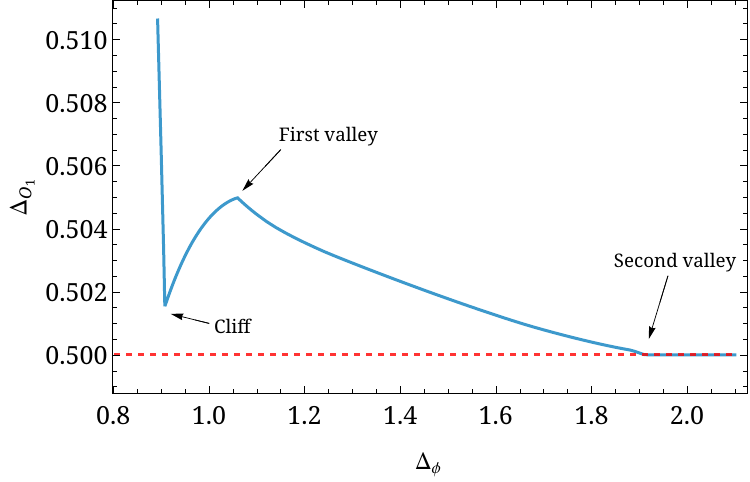}
	\caption{
		The scaling dimension of the lightest scalar operator in $d=3$ along the lower bound of the first moment, computed at $\Lambda =23$. Its anomalous dimension is locally minimized at the cliff, locally maximized at the first valley, and touches the unitarity bound at $\Delta_{\phi} \approx 1.92$, shortly before the decoupling of the second-lightest scalar. Its OPE coefficient vanishes at the same time it saturates the unitarity bound, as shown in figure~\ref{fig:second_kink_spec}. We will discuss this subtlety in section~\ref{sec:fake_primary}.
	}
	\label{fig:lightest_scalar_dim}
\end{figure}

The scalar sector around points (a) and (b) is shown in figure~\ref{fig:second_kink_spec}, where the scaling dimension of the second-lightest scalar operator shows a sudden jump. Slightly before the second-lightest scalar decouples at $\Delta_{\phi} \approx 1.95$, the scaling dimension of the lightest scalar, shown in figure~\ref{fig:lightest_scalar_dim}, reaches the unitarity bound at $\Delta_{\phi} \approx 1.92$. Numerically, we are unable to distinguish its dimension from the exact value $(d-2)/2$.

At first glance, it does not make sense to have an operator exactly at the unitarity bound, as such an operator would become a free field: its descendant $\partial^2 \mathcal{O}$ becomes null, implying an equation of motion of a free field, and the operator is expected to decouple from the spectrum. Indeed, figure~\ref{fig:second_kink_spec} shows that the OPE coefficient of the lightest scalar approaches zero as it reaches the unitarity bound. The subtlety, however, is that the moments are weighted by the \emph{product} of the squared OPE coefficient and the conformal block. While the OPE coefficient vanishes, the conformal block diverges at the unitarity bound. Their product can therefore remain finite and even dominant in the spectral density.

This is natural from the point of view of numerical bootstrap, as the semidefinite program does not directly ``see'' the divergence of the block. The bootstrap equations discussed in section \ref{numerical_bootstrap_setup} depend only on the full spectral measure, and the extremal spectra indeed show that the lightest scalar continues to contribute significantly even when its OPE coefficient tends to zero.

This phenomenon is, in fact, not new in the bootstrap literature and is known as the \emph{fake-primary effect}. In the next subsection we review this mechanism and explain how it provides a proper physical interpretation of the unitarity-saturating extremal solutions. We then apply this viewpoint to study the structure of the moment lower bounds.

\subsection{Fake-primary interpretation of lower bounds}
\label{sec:fake_primary}

Here we give a brief review of the fake-primary effect introduced in \cite{Erramilli:2020rlr, Karateev:2019pvw}, presenting the basic idea in the context of the single-correlator bootstrap.

The residue of a conformal block at any of its poles is proportional to the conformal block of its null descendant:
\begin{equation}
	g_{\Delta, \ell}(z, \bar{z}) \rightarrow \frac{R_A}{\Delta - \Delta_A^*}g_{\Delta_A, \ell_A}(z, \bar{z}) \quad \text{as } \Delta \rightarrow \Delta_A^*.
\end{equation}
Particularly, at the scalar unitarity bound\footnote{Notice that the conformal block of identical external scalars can only be singular for $\ell=0$ and $\Delta = (d-2)/2$ above or at the unitarity bound.}, the residue is given by the scalar conformal block of its shadow:
\begin{equation}
	g_{\Delta, 0}(z, \bar{z}) \rightarrow \frac{R_{\text{III}_{n=1,\ell=0}}}{\Delta - \frac{d-2}{2}} \, (4 r)^{2} \, h_{\frac{d-2}{2} + 2, 0} = \frac{R_{\text{III}_{1,0}}}{\Delta - \frac{d-2}{2}} g_{\frac{d+2}{2}, 0}(z, \bar{z}),
\end{equation}
where
\begin{equation}
	R_{\text{III}_{n=1,\ell=0}} = \frac{1}{8d} \left(\frac{d-2}{2}\right)^3.
\end{equation}
Therefore, whenever a theory contains an operator $\tilde{\mathcal{O}}$ with dimension exactly
\begin{equation}
	\tilde\Delta = \frac{d+2}{2},
\end{equation}
it can produce two distinct values for a moment (unless the moment is invariant under the shadow transformation). This ambiguity arises because the same contribution to the spectral density may be interpreted either as coming from an operator $\mathcal O$ saturating the unitarity bound, or from its shadow $\tilde{\mathcal O}$. Concrete examples of this effect also appear in the generalized free field correlators at specific values of the external weight; see appendix~\ref{appendix:fp_gff} for details.

\begin{figure}[tpb]
	\centering
	\includegraphics[width=\textwidth]{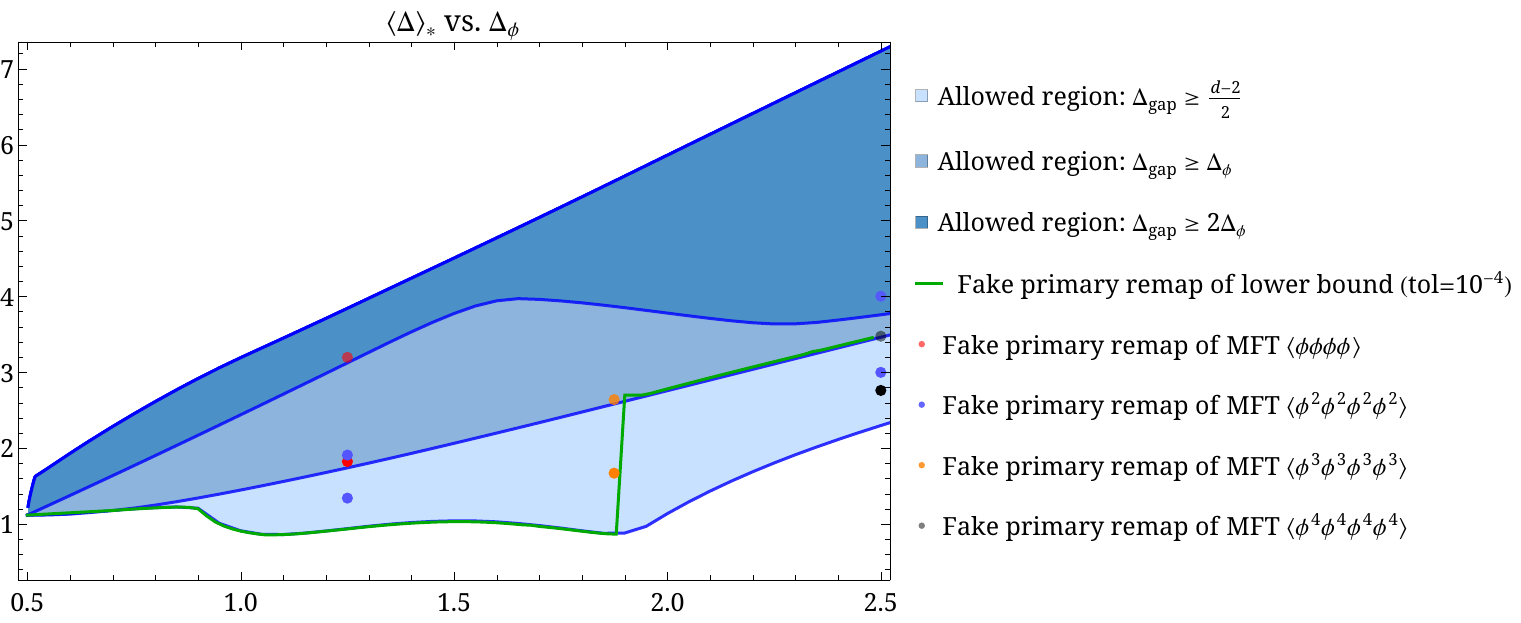}
	\caption{Fake-primary remap of the numerical lower bounds on the first moment in $d = 3$, computed at $\Lambda = 23$. The moment bounds shown here are the same as those in figure~\ref{fig:summary_3d}. We replace an operator by its shadow only if its distance from the unitarity bound is less than the tolerance. The solid green line represents the fake-primary remap of the lower bound with a tolerance of $10^{-4}$. See appendix~\ref{appendix:fp_gff} for details on the fake-primary remap of GFF moments.}
	\label{fig:fp_remap}
\end{figure}

From a numerical perspective, once an extremal solution contains a lightest scalar whose scaling dimension lies within numerical tolerance of the unitarity bound, one can perform a fake-primary remap by simply replacing this operator by its shadow while preserving its total spectral contribution. The outcome is shown in figure~\ref{fig:fp_remap}. Remarkably, after the second valley, for the region in which the extremal spectra consistently contain an operator exactly at or extremely close to the unitarity bound, the remapped moments lie nearly on top of the lower bounds obtained under the gap assumption $\Delta_{\text{gap}} \geq \Delta_\phi$.

Curiously, the external scaling dimension $\Delta_{\phi} = 1.875$, where the generalized free field OPE contains an operator exactly at the shadow of the unitarity bound, is very close to the second valley. While it remains unclear whether this alignment is accidental or inherent, the observation is compelling and deserves further investigation.

Finally, we emphasize that although the first moment can be fake-primary remapped to match the gapped lower bound, essentially following the trajectory of the $\langle \phi^n\phi^n\phi^n\phi^n \rangle$ GFF correlators at large $n$, the reconstructed higher moments do not match those of the GFF. The physical interpretation of the remapped lower bound therefore remains to be understood.

\paragraph{How gap assumptions change the lower bounds}

\begin{figure}[tpb]
	\centering
	\includegraphics[width=\textwidth]{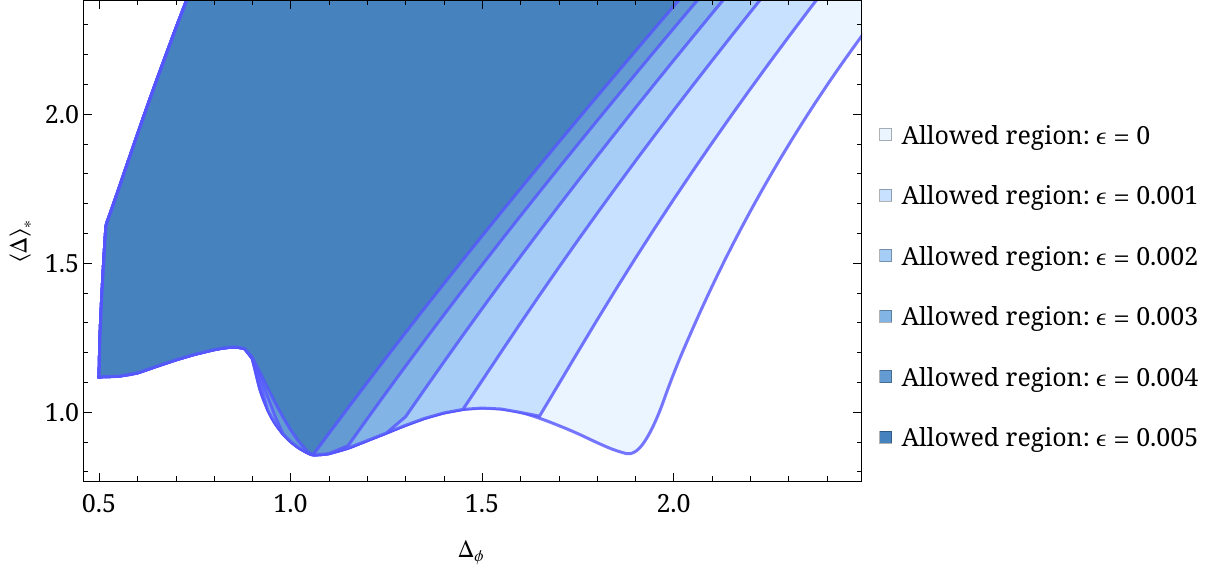}
	\caption{Upper and lower bounds on the first normalized moment $\langle \Delta \rangle_*$ with different gap assumptions $\Delta_{\text{gap}} \geq \frac{d-2}{2} + \epsilon$ imposed, computed at $\Lambda = 19$. As $\epsilon$ increases, the face of the ``cliff'' is gradually ruled out, consistent with the behavior shown in figure~\ref{fig:lightest_scalar_dim}.}
	\label{fig:gap_analysis}
\end{figure}

In \cite{Erramilli:2020rlr, Karateev:2019pvw}, varying the gap assumption was used as a diagnostic to determine whether the observed bootstrap features are genuine or are instead artifacts of the fake-primary effect. Motivated by this strategy, we perform an analogous analysis for the moment lower bounds. Figure~\ref{fig:gap_analysis} shows the upper and lower bounds on the first normalized moment $\langle \Delta \rangle_*$ in $d = 3$ as a function of the external scaling dimension while imposing a scalar gap
\begin{equation}
	\Delta_{\text{gap}} \geq \frac{d-2}{2} + \epsilon .
\end{equation}

We observe that even a mild gap assumption, $\epsilon = 0.001$, is sufficient to rule out the second valley. This behavior is expected, since the corresponding solution relies on the presence of a scalar operator saturating the unitarity bound at $\Delta = (d-2)/2$, and is therefore incompatible with a strictly positive gap. In contrast, the cliff, where the moment value undergoes its first sudden drop, persists as $\epsilon$ is increased and is not completely ruled out even for larger values of the gap. This indicates that the structures observed prior to the second valley cannot be fully attributed to the fake-primary effect. This behavior parallels observations in previous works, where lifting the scalar gap removes fake-primary features while leaving more robust bootstrap bounds only mildly affected. A more refined analysis, such as incorporating mixed-correlator constraints, may help further clarify the origin of this feature, and we leave this direction for future work.

\paragraph{Deformation across dimensions}

As we have seen throughout this work, the lower bounds on the moments generically exhibit two families of kinks, both associated with the decoupling of low-lying scalar operators. The second valley, in particular, marks the saturation of the scalar unitarity bound, beyond which the numerical solution can be consistently reinterpreted through a fake-primary remap involving a scalar at $\Delta=\frac{d+2}{2}$. The detailed realization of this mechanism, however, depends on the spatial dimension.

In exactly $d=2$, as shown in figure~\ref{fig:summary_2d}, the lower bound exhibits no kinks of the type observed in higher dimensions and instead closely follows the linear trajectory $\sqrt{2}\,\Delta_\phi$, coinciding with the bound obtained by imposing $\Delta_{\text{gap}} \geq \Delta_\phi$. This behavior changes \emph{discontinuously} when moving slightly away from two dimensions. 

\begin{figure}[tbp]
	\centering
	\includegraphics[width=0.9\textwidth]{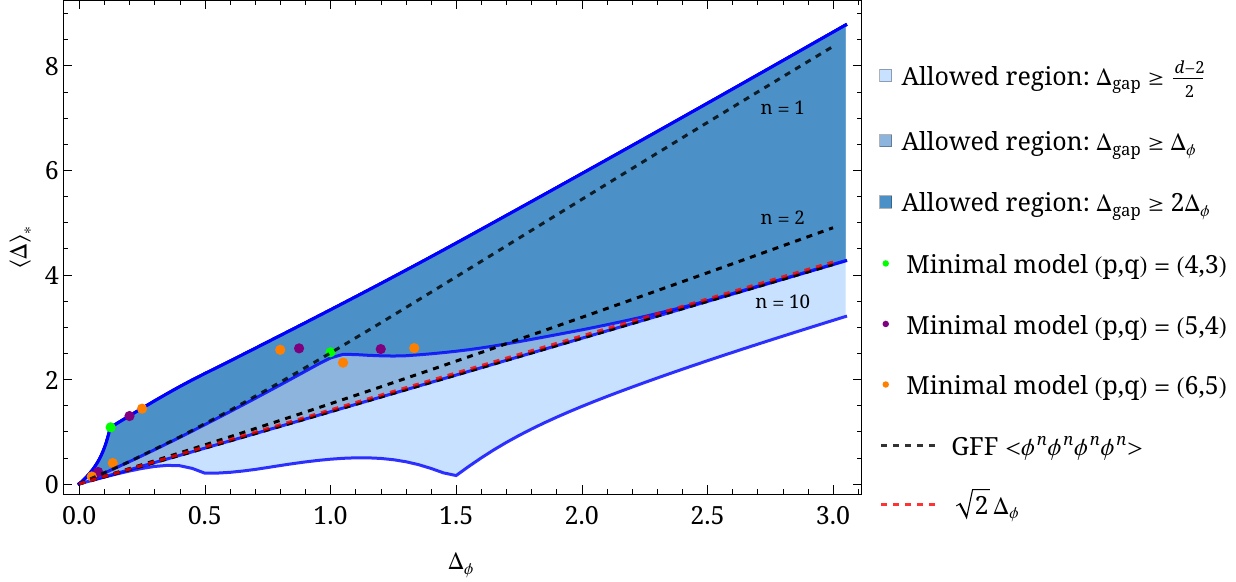}
	\caption{Moment bounds in $d = 2 + 10^{-6}$, computed at $\Lambda = 19$. With a unitary gap assumption, the lower bound exhibits two sharp kinks at the simple values $\Delta_{\phi} = \tfrac{1}{2}$ and $\Delta_{\phi} = \tfrac{3}{2}$. After a fake-primary remap (see section~\ref{sec:fake_primary}), the entire curve is deformed upward to coincide with the lower bound obtained by imposing $\Delta_{\text{gap}} \geq \Delta_{\phi}$. Although this behavior is fake-primary driven in the limit $d \rightarrow 2^+$, in general dimensions $2 < d < 5$ the scalar operator before the second valley lies strictly above the unitarity bound, and the associated lower bounds are not entirely governed by fake-primary effects. Appendix~\ref{appendix:minimal_model} describes our method for computing the minimal model moments.
	}
	\label{fig:2_plus_epsilon}
\end{figure}

In the special limit $d\to 2^+$, distinct geometric landmarks that are well separated at generic dimensions merge into a single kink, where operator decoupling coincides with a local minimum of the moment. In $d=2+10^{-6}$ the lower bound develops two sharp kinks at the half-integer values $\Delta_\phi=\tfrac12$ and $\tfrac32$, as shown in figure~\ref{fig:2_plus_epsilon}. Operator decoupling occurs exactly at these points, and the entire lower-bound contains a scalar saturating the unitarity bound. Consequently, the trajectory is ``fake'' in the sense that, after an appropriate fake-primary remapping, it coincides numerically with the lower bound obtained by imposing $\Delta_{\text{gap}} \geq \Delta_\phi$. Imposing a tiny gap immediately collapses the bound onto the same curve. This trajectory asymptotically matches that of generalized free field correlators $\langle \phi^n\phi^n\phi^n\phi^n\rangle$ at large $n$ and aligns with the bounds observed in strictly $d=2$. For comparison, we also include the moments of known theories, including the low-lying minimal models and the generalized free field theory. 

The origin of the discontinuity can be traced to the behavior of the scalar conformal block at the unitarity bound, $\frac{d-2}{2}$. In $d=2+\epsilon$, the conformal block diverges as the scalar dimension approaches $\epsilon/2$, so placing an operator exactly at the unitarity bound, interpreted as the fake-primary image of another solution containing a scalar at the location of its shadow, leads to a significant reduction of the moment. In contrast, in exactly $d=2$, the global scalar block is regular at the unitarity bound, preventing an analogous mechanism. 

Finally, we emphasize that in $d=2+10^{-6}$, although the first moment after the second valley can be fake-primary remapped to match that of the GFF, the extremal spectra themselves do not coincide with those of the GFF. This discrepancy appears in the higher moments, indicating that the underlying solutions are not genuine GFFs. The physical interpretation of these extremal solutions therefore remains unclear and merits further investigation.

\subsection{The $\Delta_{\text{gap}} \geq 2\Delta_\phi$ lower bound}
\label{subsec:gap_2phi}

\begin{figure}[tbp]
	\centering
	\includegraphics[width=\textwidth]{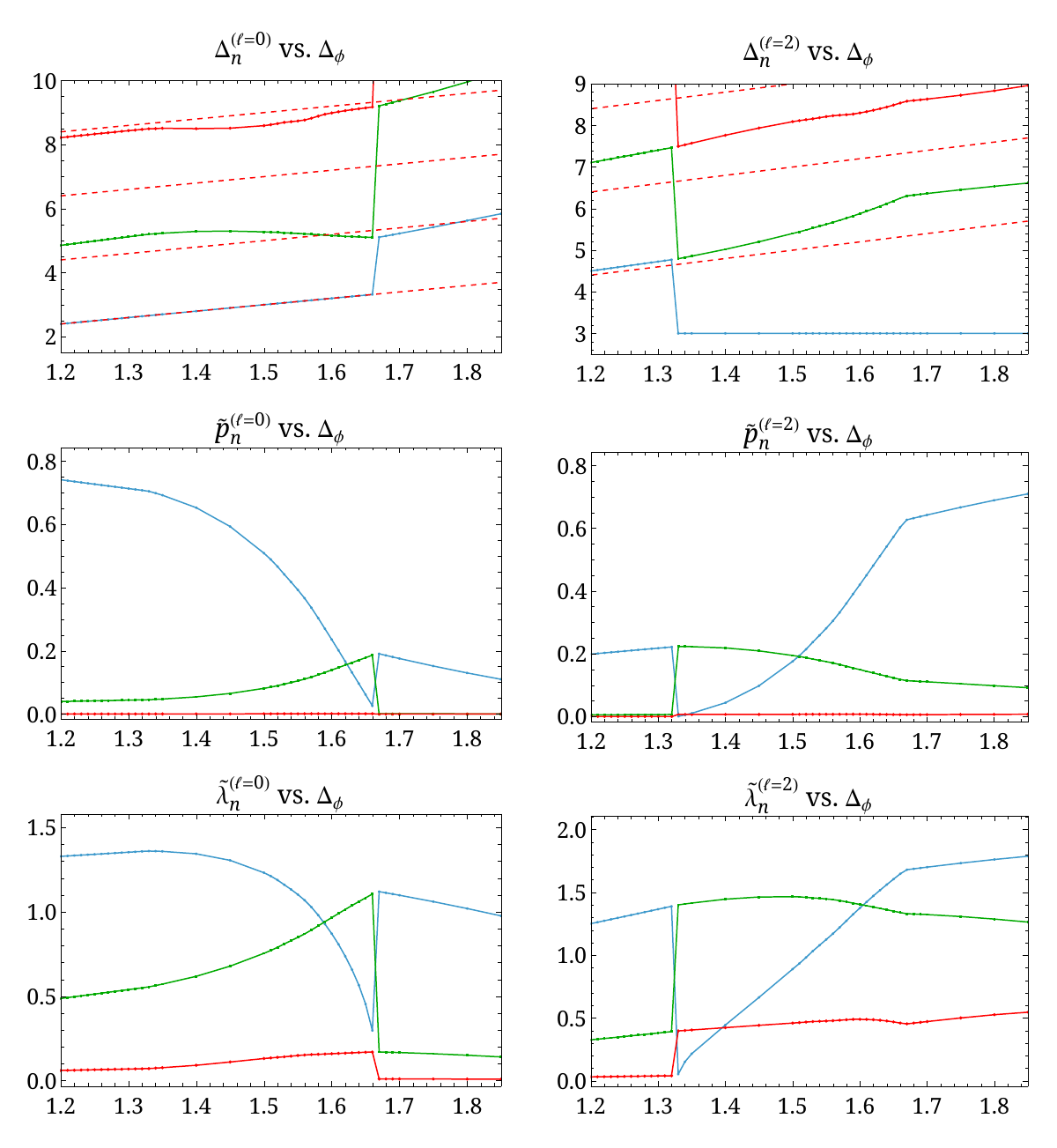}
	\caption{
		Extremal spectra of the leading scalar and spin-2 operators along the $\Delta_{\text{gap}} \geq 2\Delta_{\phi}$ lower bound in $d=3$, computed at $\Lambda = 23$. The red dashed lines represent $2\Delta_{\phi} + 2n$ for $n=0,1,2,3$. The stress tensor enters the spectrum at $\Delta_{\phi} \approx 1.3$. Its contribution to the spectral density then increases steadily and eventually acquires an anomalous dimension at $\Delta_{\phi}\approx 3.5$. At $\Delta_{\phi}\approx 1.63$, where the gapped lower bound reaches its maximal value, the lightest scalar operator decouples. Prior to this point, its scaling dimension remains fixed at the assumed gap, $2\Delta_{\phi}$.
	}
	\label{fig:gap_2phi_spec_3D}
\end{figure}

We end this section by highlighting the structures of the lower bound obtained under the gap assumption $\Delta_{\text{gap}} \geq 2\Delta_\phi$.
As a function of the external scaling dimension $\Delta_\phi$, the extremal solutions interpolate between three qualitatively distinct regimes:
\begin{enumerate}[label=(\alph*)]
	\item a family of solutions that trace the generalized free field (GFF) $\langle \phi\phi\phi\phi\rangle$ correlators at small $\Delta_\phi$,
	\item an intermediate regime where the moment bound reaches a local maximum, accompanied by the decoupling of the scalar at the gap $2\Delta_\phi$,
	\item and a final regime approximately saturated by the GFF correlators $\langle \phi^n \phi^n \phi^n \phi^n \rangle$ at large $n$.
\end{enumerate}
This interpolating behavior is observed consistently across different spatial dimensions, as shown in figure~\ref{fig:summary_all}.

For light correlators, the gap assumption significantly shrinks the allowed theory space, and the lower bound is tightly saturated by the generalized free field $\langle \phi\phi\phi\phi\rangle$ trajectory. Indeed, the extremal spectra approximately reproduce the constant-twist double-trace operators. In particular, the lightest scalar operator exactly follows $2\Delta_\phi$, saturating the gap assumption.

As $\Delta_\phi$ increases, the anomalous dimensions of operators near the constant-twist trajectories gradually develop, and at the local maximum of the lower bound, the scalar operator at $2\Delta_\phi$ decouples from the extremal spectrum. This scalar decoupling occurs across various spatial dimensions and remains very close to the local maximum of the lower bound.  Figure~\ref{fig:gap_2phi_spec_3D} shows the extremal spectra along the gapped lower bound in a representative slice, $d=3$.

Beyond this point, the lower bound decreases and asymptotically approaches the linear trajectory $\sqrt{2}\,\Delta_\phi$. This behavior coincides with the $\Delta_{\text{gap}} \geq \Delta_\phi$ lower bound, or with the first moment of the GFF correlators $\langle \phi^n \phi^n \phi^n \phi^n \rangle$ at large $n$.

\paragraph{Deformation across dimensions}
We extend $\Delta_\phi$ to larger values in this gapped setup to track the approach to the asymptotic linear trajectory. For $d \lesssim 5$, there exists an intermediate window along the gapped lower bound in which the extremal spectra contain a conserved stress tensor. Intriguingly, we observe that the stress tensor enters the extremal spectrum when the external scaling dimension is close to $\frac{d+2}{4}$, where the fake-primary remap of the GFF becomes first possible, as argued in appendix~\ref{appendix:fp_gff}. As $\Delta_\phi$ increases further, the stress tensor decouples as the solution merges into the asymptotic linear regime. It would be interesting to better understand this interpolating behavior analytically and how it may be related to known physical theories.


\section{Discussion}
\label{sec:discussion}


The numerical moment bootstrap developed in this work extends our understanding of OPE moment variables into the light correlator regime. Through the use of semidefinite programming, it provides numerically rigorous bounds that complement the analytic power-law behavior derived in the heavy correlator limit. Assuming maximal Shannon entropy of the OPE spectrum, the extremal solutions allow for accurate reconstruction of the associated spectral density, confirming convergence toward generalized free field theories while yielding systematic corrections to the spectral density. In this way, the numerical moment bootstrap establishes a concrete bridge between numerical bootstrap techniques and analytical heavy limit analysis.

Conceptually, the moment formulation offers a complementary perspective on the conformal bootstrap by probing the spectrum in a global and coarse-grained manner, compressing detailed operator information into weighted averages that remain well defined even when the spectrum becomes dense. Perhaps surprisingly, such collective observables can nevertheless be studied in a numerically rigorous way. Beyond the single heavy correlators studied in this work, the moment formulation naturally extends to mixed correlators involving both heavy and light external operators. In particular, heavy-heavy-light-light correlators provide a promising arena where moment variables may probe finite-temperature physics and aspects of thermalization in conformal field theories. We leave a systematic exploration of such mixed systems to future work.


Another notable advantage of this framework is its ability to unify a wide range of familiar bootstrap phenomena. Low-lying moments of scaling dimension and spin naturally reproduce the Ising-model kinks and encode gap and central-charge extremization within a single optimization problem. Compared to traditional gap-exclusion methods, which require repeated scans over the parameter space, the moment approach identifies physically relevant theories more efficiently while simultaneously yielding globally rigorous bounds on the operator spectrum. This points to a natural future direction: by bounding moment variables using mixed-correlator constraints over the bootstrap island, one may establish rigorous statements about conformal field theories not only along the boundary of the island but also in its interior \cite{mixed-moment-bootstrap:wip}.


Most intriguingly, the numerical moment bootstrap reveals previously unexplored geometric structures on the other side of the bootstrap landscape. The lower bounds of moment variables exhibit robust features across spatial dimensions, closely tied to nontrivial spectral events including operator decoupling, anomalous-dimension extremization, and the fake-primary effect. In particular, the emergence of two continuous families of kinks across $2<d<6$ demonstrates that moment variables are sensitive to spectral reorganizations that are otherwise difficult to access within traditional bootstrap setups.

These geometric features admit a simple organization in terms of lightest scalar operators. The ``pond'' region is controlled by the decoupling and reappearance of the lightest scalar operator. As its OPE coefficient is numerically small and its imprint on the moment geometry is therefore subtle, it will be important to understand whether this structure sharpens or persists as the derivative order is increased.

The two sharp kinks across dimensions are closely tied to the second-lightest operator, whose scaling dimension closely follows $2\Delta_{\phi}$. The first kink (the ``cliff''), which is continuously connected to the free scalar theory in $d = 6$, features the decoupling of this operator. At the second kink, the operator reappears, while the lightest scalar approaches the unitarity bound, and the extremal spectra simultaneously minimize multiple low-lying moments. Whether these structures arise from renormalization group flows of known interacting scalar theories, or instead probe an approximately unitary sector of more exotic solutions, remains an open question. Clarifying the physical origin of the ``cliff'' will likely require additional input, for instance from mixed-correlator bootstrap analyses that may isolate these solutions into bootstrap islands. Determining the ultimate fate of the second kink in higher spatial dimensions likewise remains an intriguing open direction.

Between the two kinks lies the ``hill'' region, where the moment-minimizing solutions probe qualitatively different directions in the theory space, with the reconstructed correlators showing unbounded growth. In this regime, the OPE is dominated by only a few low-lying operators, suggesting that analytic tools such as the Lorentzian inversion formula~\cite{Caron-Huot:2017vep} may provide further insight into the observed spectra. Finally, the gapped lower bound exhibits an interpolation between asymptotic regimes with an intermediate window of local theories, further motivating analytic investigations of these structures.

In this work, we defined moments at the self-dual point $z = \zb = 1/2$, a kinematic configuration where conformal blocks suppress contributions from operators with large scaling dimensions. More generally, we can define moments of the OPE in different kinematic regimes, which would be sensitive to qualitatively different sets of operators. For example, if we study moments of the OPE sum weighted by conformal blocks evaluated near the double light cone limits, the resulting quantities would naturally be most sensitive to the operators appearing in the leading twist trajectory. Alternatively, if we study moments of the OPE deep in the Regge limit, these quantities may be used to characterize Regge trajectories, which play an important role in the dynamics of this regime. It would be interesting to understand which moments are naturally adapted to different kinematic limits, and how they may be bounded numerically or computed perturbatively in expansions around these limits.


In conclusion, the moment bootstrap reveals both subtle and unexpected geometric structures in the space of CFT four-point functions of identical scalars, while opening up several promising directions for future work, including numerically probing the heavy correlator regime and reinforcing existing precision studies of conformal data. We hope that this work offers a clear map of the emerging ``moment landscape'' and encourages further exploration of its physical origins.

\acknowledgments

We thank Wei Li for numerous discussions and collaboration on related CFT moment problems, Liam Fitzpatrick, Yu-tin Huang, Shao-Cheng Lee, Tony Liu, Matthew Mitchell, and Balt van Rees for additional discussions, and Yuan Xin for valuable guidance on computational aspects in the early stages of this work. The authors were supported by DOE grant DE-SC0017660. Computations were performed on the Yale Grace computing cluster, supported by the facilities and staff of the Yale University Faculty of Sciences High Performance Computing Center.

\appendix

\section{Rational approximation of conformal blocks}
\label{appendix:rational_approx}

Our setup for the numerical bootstrap starts from the rational approximation of conformal blocks. Extensively used in numerical bootstrap studies, it expresses a conformal block in terms of a rational function of the scaling dimension $\Delta$ of the operator exchanged. This approximation is based on an expansion in the radial coordinate $r$, related to the cross ratio $z$ via \cite{Hogervorst:2013sma}
\begin{equation}
	\rho = r e^{i \theta} = \frac{1-\sqrt{1-z}}{1 + \sqrt{1-z}},
\end{equation}
and $\eta = \cos{\theta}$. The idea is that the conformal block develops singularities at non-physical scaling dimensions $\Delta$, which gives a recursion relation of the conformal block in terms of a sum over poles in $\Delta$. Such a relation was first discovered by Zamolodchikov \cite{Zamolodchikov_1984,Zamolodchikov_1987} in 2d and was generalized to higher dimensions in \cite{Kos:2013tga} for identical external scalar primaries. Here, we follow the convention in \cite{Poland:2018epd} and write the conformal block as
\begin{equation}
	g_{\Delta, \ell}(r, \eta) = (4r)^\Delta h_{\Delta, \ell}(r, \eta),
\end{equation}
where the regularized conformal block $h_{\Delta, \ell}(r, \eta)$ admits the recursive representation
\begin{equation}
	h_{\Delta, \ell}(r, \eta) = h_{\infty, \ell}(r, \eta) + \sum_{A} \frac{R_A}{\Delta - \Delta_A^*}(4r)^{n_A} h_{\Delta_A^* + n_A, \ell_A}(r, \eta),
\end{equation}
with the locations of the poles $\Delta_A^*$ and the residues $R_A$ given in table~\ref{recursion_pol_and_res}.

\begin{table}[tpb]
	\centering
	\renewcommand{\arraystretch}{1.4}
	\setlength{\tabcolsep}{4pt}
	\begin{tabular}{| c | c | c | c | c |}
		\hline
		$A$ & $\Delta_A^*$ & $n_A$ & $\ell_A$ & $R_A$ \\
		\hline
		\rule{0pt}{3.2ex}\rule[-1.6ex]{0pt}{0pt}$\text{I}_n (n \in 2\mathbb{N})$  & $1-\ell-n$      & $n$   & $\ell+n$ &
		{\footnotesize
			$\begin{aligned}
				R_{\mathrm{I}_n}
				& = \frac{-n (-2)^n}{(n!)^2}
				\left(\frac{1-n}{2}\right)_n^2
			\end{aligned}$
		}
		\\
		\hline
		\rule{0pt}{3.2ex}\rule[-1.6ex]{0pt}{0pt}$\text{II}_n (n \in 2\mathbb{N})$ & $\ell+d-1-n$    & $n$   & $\ell-n$ &
		{\footnotesize
			$\begin{aligned}
				R_{\mathrm{II}_n}
				 & = \frac{-n\,\ell!}{(-2)^n (n!)^2 (\ell-n)!} \\
				 & \qquad\times
				\frac{(d+\ell-n-2)_n}{
					\left(\frac{d}{2}+\ell-n\right)_n
					\left(\frac{d}{2}+\ell-n-1\right)_n}
				\left(\frac{1-n}{2}\right)_n^2
			\end{aligned}$
		}
		\\
		\hline
		\rule{0pt}{3.2ex}\rule[-1.6ex]{0pt}{0pt}$\text{III}_n (n \in \mathbb{N})$ & $\frac{d}{2}-n$ & $2n$  & $\ell$   &
		{\footnotesize
			$\begin{aligned}
				R_{\mathrm{III}_n}
				 & = \frac{-n (-1)^n
					\left(\frac{d}{2}-n-1\right)_{2n}}
				{(n!)^2
					\left(\frac{d}{2}+\ell-n-1\right)_{2n}
					\left(\frac{d}{2}+\ell-n\right)_{2n}}
			\end{aligned}$
			}
		\\
		\hline
	\end{tabular}
	\caption{The locations of the poles of the conformal block and the corresponding residue data, adapted from \cite{Poland:2018epd}.}
	\label{recursion_pol_and_res}
\end{table}

Therefore, by iterating the recursion relation up to order $r_{\max}$ in the radial coordinate $r$, we arrive at the following approximation for the conformal block and its derivatives,
\begin{equation}
	\partial_z^m \partial_{\bar{z}}^n g_{\Delta, \ell}(1/2,1/2) \approx \frac{(4r_*)^\Delta}{\prod_{A}(\Delta - \Delta_A^*)} P^{mn}_\ell(\Delta),
\end{equation}
at the crossing-symmetric point $r_* = 3 - 2 \sqrt{2}, \, \eta = 1$. For maximized computational efficiency, we would like the degree of the polynomial $P_\ell^{m,n}(\Delta)$ to be as small as possible. Following the method first proposed in \cite{Kos:2013tga}, a two-sided Padé approximation is used to reduce the number of poles in the rational function, keeping those that appear below order $\kappa$ in the radial expansion while matching the first $\lfloor \frac{|P_\kappa| + 1}{2} \rfloor$ derivatives at the unitarity bound and the remaining $\lfloor \frac{|P_\kappa|}{2} \rfloor$ at $\Delta \rightarrow \infty$, where $|P_\kappa|$ denotes the number of poles appearing at radial expansion order $\kappa$.

All conformal block derivatives used in this work were generated using a custom Julia library, \texttt{blockDeriv.jl}~\cite{blockDeriv}, developed for efficiently computing derivatives of scalar conformal blocks with identical external operators, including kinematic configurations away from the self-dual point. In exactly $d=2$, we instead generated the derivatives using \texttt{scalar\_blocks}~\cite{scalarblocks} for its specialized treatment of the pole structure in even dimensions. The \texttt{blockDeriv.jl} package is currently being prepared for public release.

Unless stated otherwise, throughout this work we adopt the truncation parameters specified in table~\ref{tab:param}.
The duality gap is always chosen to be $10^{-30}$, except in regions where the extremal spectra require a tighter gap to stabilize, in which case it is set to $10^{-60}$. We have verified that the moment bounds are insensitive to the truncation parameters.

\begin{table}[H]
	\centering
	\begin{tabular}{cccc}
		\hline
		$\Lambda$ & $\kappa$ & $\ell_{\max}$ & $r_{\max}$ \\
		\hline
		5         & 8        & 30            & 60         \\
		9         & 8        & 30            & 60         \\
		11        & 8        & 30            & 60         \\
		13        & 8        & 30            & 60         \\
		19        & 14       & 50            & 60         \\
		23        & 18       & 50            & 60         \\
		27        & 20       & 50            & 60         \\
		\hline
	\end{tabular}
	\caption{Truncation parameters for bootstrap computations in this work.}
	\label{tab:param}
\end{table}

\section{Moments from 2d correlators}
\label{appendix:minimal_model}

Minimal models play a central role in the conformal bootstrap. They were among the first theories studied using bootstrap techniques and provide exactly solvable examples with significant implications. Computing moment variables in these CFTs is therefore a natural and important task: not only is their OPE data explicitly known, but they also serve as a sanity check for locating exactly solvable theories within the moment landscape charted in this work.

In this appendix, we describe how the minimal-model and vertex operator moments shown in figures~\ref{fig:summary_2d} and~\ref{fig:2_plus_epsilon} are computed. The method is conceptually straightforward. The unitary minimal models are labeled by $(p,q)=(p,p+1)$ with $p \geq 3$, and the conformal dimensions of Virasoro primary operators are given by
\begin{equation}
	h_{r,s}=\frac{(ps-qr)^2-(p-q)^2}{4pq},
\end{equation}
where the positive integers $r$ and $s$ lie in the ranges
\begin{equation}
	1\leq r\leq p - 1, \qquad 1\leq s\leq p.
\end{equation}
The structure constants of these theories are known in closed form, following the classic work of Dotsenko and Fateev~\cite{Dotsenko:1984nm, Dotsenko:1984ad, Dotsenko:1985hi}. In principle, this makes the computation of correlation functions entirely explicit once the Virasoro conformal blocks are available. However, since our goal is to compute moments defined as weighted averages over global conformal primaries, it is necessary to decompose Virasoro blocks into sums of global blocks.

Fortunately, as emphasized in~\cite{Perlmutter:2015iya}, several useful representations of Virasoro blocks can be obtained from the recursion relations discovered by Zamolodchikov~\cite{Zamolodchikov_1984,Zamolodchikov_1987}, including expansions in terms of an infinite sum of global hypergeometric blocks. We make use of this representation to express the Virasoro blocks as sums over the global conformal blocks, up to a truncation $\Delta_{\max}$ in the scaling dimensions. Combined with the known structure constants, this determines the global conformal data, which enables us to compute the moments directly from their definition. We present their values in table~\ref{tab:minimal_model_moments}, computed with $\Delta_{\max} = 14$. We have checked that the resulting values are stable under increases in $\Delta_{\max}$ and are quoted to three decimal places.

\begin{table}[H]
	\centering
	\renewcommand{\arraystretch}{1.15}
	\setlength{\tabcolsep}{6pt}

	\begin{tabular}{c c c c c}
		\hline
		$(p,q)$
		 & $(r,s)$ & $h_{r,s}$        & $\Delta_\phi$    & $\langle \Delta \rangle_\star$ \\
		\hline
		\multirow{2}{*}{\textbf{(3, 4)}}
		 & $(1,2)$ & $\tfrac{1}{16}$  & $\tfrac{1}{8}$   & $1.081$                        \\
		 & $(1,3)$ & $\tfrac{1}{2}$   & $1$              & $2.516$                        \\
		\hline
		\multirow{4}{*}{\textbf{(4, 5)}}
		 & $(1,2)$ & $\tfrac{1}{10}$  & $\tfrac{1}{5}$   & $1.297$                        \\
		 & $(1,3)$ & $\tfrac{3}{5}$   & $\tfrac{6}{5}$   & $2.578$                        \\
		 & $(2,1)$ & $\tfrac{7}{16}$  & $\tfrac{7}{8}$   & $2.592$                        \\
		 & $(2,2)$ & $\tfrac{3}{80}$  & $\tfrac{3}{40}$  & $0.228$                        \\
		\hline
		\multirow{6}{*}{\textbf{(5, 6)}}
		 & $(1,2)$ & $\tfrac{1}{8}$   & $\tfrac{1}{4}$   & $1.439$                        \\
		 & $(1,3)$ & $\tfrac{2}{3}$   & $\tfrac{4}{3}$   & $2.598$                        \\
		 & $(2,1)$ & $\tfrac{2}{5}$   & $\tfrac{4}{5}$   & $2.564$                        \\
		 & $(2,2)$ & $\tfrac{1}{40}$  & $\tfrac{1}{20}$  & $0.141$                        \\
		 & $(2,3)$ & $\tfrac{1}{15}$  & $\tfrac{2}{15}$  & $0.398$                        \\
		 & $(2,4)$ & $\tfrac{21}{40}$ & $\tfrac{21}{20}$ & $2.319$                        \\
		\hline
	\end{tabular}
	\caption{The first normalized moment variable for identical external scalars, with the identity contribution removed, in the low-lying minimal models. Here we list the moments appearing in figure~\ref{fig:2_plus_epsilon}, corresponding to correlators of relevant operators.}
	\label{tab:minimal_model_moments}
\end{table}

The vertex operator moments can be computed in a similar way, now by decomposing the correlator into a positive sum of global conformal blocks. For the identical scalar correlator setup considered in this work, we take the cosine vertex operator
\begin{equation}
	C_{\alpha} = \frac{V_{\alpha} + V_{-\alpha}}{\sqrt{2}},
\end{equation}
which gives a non-vanishing four-point function
\begin{equation}
	G(z,\bar z) = \tfrac12 \left(v^{\phi} + u^{2\phi} v^{-\phi} + v^{-\phi}\right).
\end{equation}
We expand the correlation function in the $s$-channel OPE limit $z = \zb = 0$ up to order $z_{\max}$, so that only operators with sufficiently small $\Delta$ contribute at this order. We then match this expansion to a sum over global conformal blocks to extract the OPE data. The operators appearing in the OPE consist of the global primaries in the Verma modules of the vacuum and of the vertex operator $C_{2\alpha}$. Their quantum numbers therefore organize into two families, spaced by integers above the corresponding Virasoro primaries. With $z_{\max}=8$, we compute the moments shown in figure~\ref{fig:summary_all}.

\section{Ising moments across dimensions}
\label{appendix:Ising_across_dims}

The seminal work of Wilson and Fisher \cite{Wilson:1971dc} first established that the critical point of the Ising model corresponds to a nontrivial infrared fixed point of the $\phi^4$ theory below four dimensions, unifying the 2d and 3d Ising critical points with the Gaussian theory in $d = 4$. The conformal bootstrap has provided precise nonperturbative evidence for this picture. In particular, \cite{El-Showk:2013nia} extended the numerical bootstrap to fractional spatial dimensions, demonstrating the existence of a smooth family of CFTs interpolating between the 2d and 3d Ising models and the free theory in 4d. More recently, refined analyses have mapped this family with further accuracy, extracting the critical exponents \cite{Bonanno:2022ztf}.

\begin{figure}[tbp]
	\centering
	\includegraphics[width=0.9\textwidth]{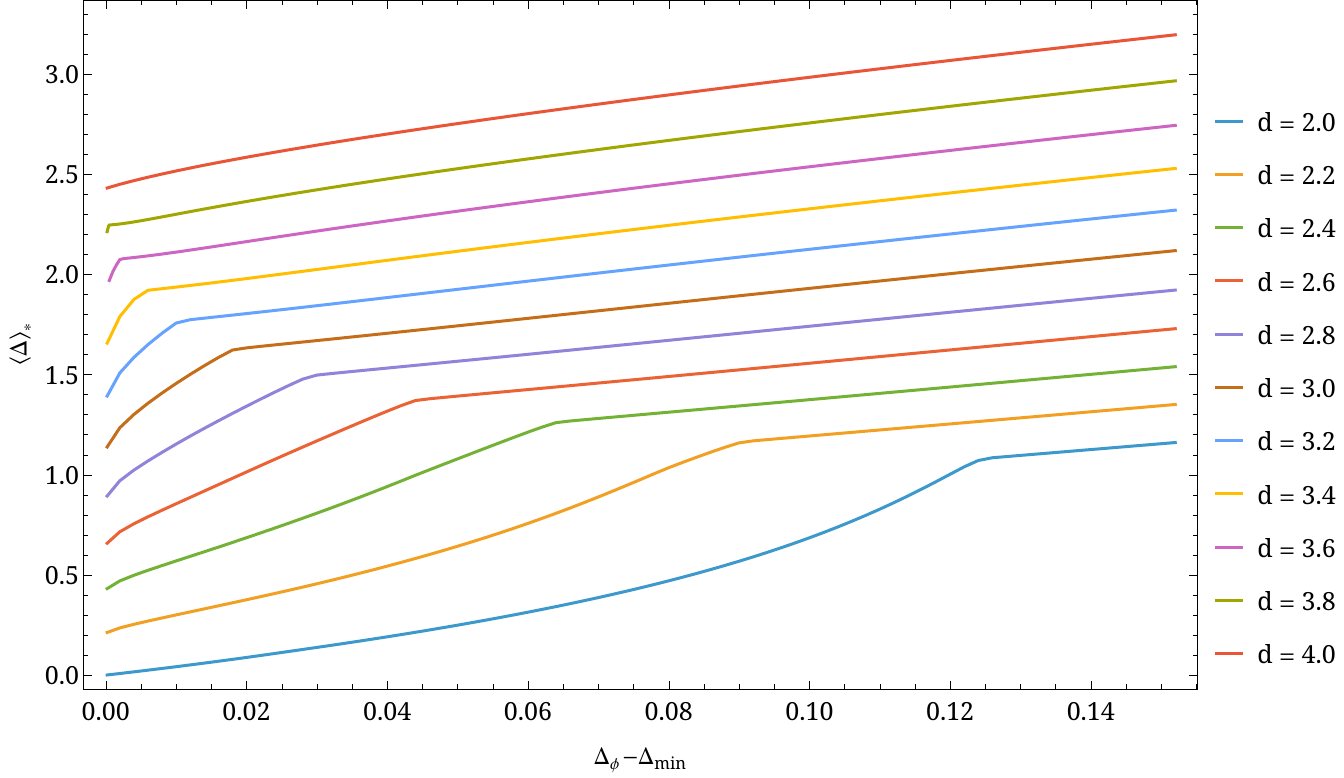}
	\caption{Upper bounds on the first normalized moment (with the identity operator excluded), plotted as a function of the external scaling dimension $\Delta_\phi$ after subtracting the unitarity bound $\Delta_{\min} = \frac{d-2}{2}$, in various spatial dimensions $d$. These bounds were computed at $\Lambda = 19$. One can see here the evolution of the Ising kinks across different dimensions.}
	\label{fig:ising_across_dim}
\end{figure}

In our framework, the kinks associated with the Ising fixed points can be efficiently tracked across spatial dimensions using the moment variables discussed in this work. Figure~\ref{fig:ising_across_dim} shows the first normalized moment (with the identity contribution removed) as a function of the external dimension $\Delta_\phi$, after subtracting the unitarity bound $\Delta_{\min} = \frac{d-2}{2}$.

\section{Maximum entropy reconstruction of coarse-grained spectra}
\label{appendix:max_entropy}

The maximum-entropy principle provides a natural prescription for selecting a probability measure when only partial information is available. Originally formulated in the context of statistical mechanics by Jaynes~\cite{Jaynes:1957zza}, it states that among all distributions consistent with a given set of constraints, the one maximizing the Shannon entropy constitutes the least biased choice. Its first systematic applications to quantum-mechanical systems were carried out by Mead and Papanicolaou~\cite{Mead_Papanicolaou_1984}, who employed the method to reconstruct densities of states in harmonic solids and dynamical correlation functions in quantum spin systems. In quantum field theory, maximum-entropy techniques have become a standard tool for reconstructing spectral functions from a finite set of Euclidean correlators, particularly in lattice QCD studies~\cite{Asakawa:2000tr}. Related ideas have also appeared in other areas of high-energy physics, for instance in the reconstruction of parton distribution functions from a finite set of moments~\cite{Zhang:2023oja}.

Here we outline the method of reconstructing a spectral density from a finite number of moments. Given the first few normalized moments,
\begin{equation}
	\langle \Delta^k \rangle_* = \int d\Delta \, \tilde{p}(\Delta) \Delta^k, \quad k = 1, 2, ..., N,
\end{equation}
the task is to determine a positive distribution $\tilde p(\Delta)$ that reproduces these values. For general values of the moments, the solution to the moment problem is not unique, as infinitely many distributions can share the same first $N$ moments. The max-entropy method provides a natural criterion for selecting among them: it chooses the distribution that maximizes the Shannon entropy
\begin{equation}
	S[\tilde p] = - \int d \Delta \, \tilde p(\Delta) \, \log \tilde p (\Delta).
\end{equation}
By the method of Lagrange multipliers, the maximizing distribution always takes the exponential form
\begin{equation}
	\tilde p(\Delta) = \frac{1}{Z(\lambda)} \exp \left( -\sum_{n=1}^N \lambda_n \Delta^n \right),
\end{equation}
where the coefficients $\lambda_n$ are determined by enforcing the conditions that the spectral density reproduces the first $N$ given moments. In practice, these parameters can be efficiently solved numerically using existing implementations such as the Python package \href{https://github.com/saadgroup/PyMaxEnt}{\texttt{PyMaxEnt}} \cite{SAAD2019100353}.

In the context of this work, the max-entropy reconstruction is used in section \ref{sec:heavy correlator} to obtain a smooth, positive approximation to the discrete OPE spectral measure, providing a weight-interpolating function that captures the collective behavior of heavy operators.

\section{Fake-primary remap of generalized free field correlators}
\label{appendix:fp_gff}

As argued in section \ref{sec:new-kinks}, when a theory contains a scalar operator exactly at $\Delta = \frac{d+2}{2}$, the fake-primary effect allows us to remap the operator to its shadow at the unitarity bound while maintaining crossing and unitarity. A concrete example is provided by the generalized free field theory: let us consider the correlation function $\langle\phi \phi \phi \phi\rangle$, where the exchanged operators in the $\phi \times \phi$ OPE have scaling dimensions
\begin{equation}
	\Delta_{[\phi\phi]_{n,\ell}} = 2\Delta_\phi + 2n + \ell.
\end{equation}
Therefore, the scenario above happens only when
\begin{equation}
	2 \Delta_{\phi} + 2n + \ell = \frac{d+2}{2}, \quad \Delta_\phi \geq \frac{d-2}{2},
\end{equation}
which is possible only when $n = \ell = 0$, and we find
\begin{equation}
	\Delta_\phi = \frac{d + 2}{4}, \quad d \leq 6.
\end{equation}

We can extend this analysis to $\langle \phi^K \phi^K \phi^K \phi^K \rangle$, where the scaling dimensions of the exchanged scalar operators are
\begin{equation}
	\Delta_{n,k,l} = 2 k \Delta_\phi + 2n + \ell,
\end{equation}
where $k = 1,\dots,K$. Writing the composite operator as $\Phi = \phi^K$, there is an operator with scaling dimension exactly $\frac{d+2}{2}$ only when $n = \ell = 0$, which gives us
\begin{equation}
	\Delta_{\Phi} = \frac{K}{k} \frac{d+2}{4}, \quad d \leq \frac{4k+2}{2k-1}.
\end{equation}
Consequently, the criteria on $k$ and $d$ are given by
\begin{equation}
	d \leq 6, \quad k \leq \frac{d+2}{2d-4}.
\end{equation}
In $d = 3$, only $k = 1, 2$ are allowed, which gives us the lowest external scaling dimensions $\Delta_{\Phi} = 1.25, 1.875, 2.5, ...$, exactly the points labeled in figure~\ref{fig:fp_remap}.

\bibliographystyle{JHEP}
\bibliography{biblio.bib}

\end{document}